%% file: main.tex
\documentclass[12pt]{article}

\usepackage[margin=1in]{geometry}
\usepackage{xcolor}
\usepackage{graphicx}
\usepackage{epsfig}
\graphicspath{{./eps/}}
\usepackage{epstopdf}
\usepackage{subfig}

\usepackage{amsmath}
\usepackage{amssymb}
\usepackage{amsthm}

\usepackage{rotating}
\usepackage{verbatim}
\usepackage{hyperref}
\usepackage{bm}
\usepackage[normalem]{ulem}
\usepackage[authoryear, semicolon]{natbib}
\usepackage{bbm}
\usepackage{array}
\usepackage{enumerate}
\usepackage{float}
\usepackage{authblk}
\usepackage{algorithm,algorithmic}

\usepackage[flushleft]{threeparttable}
\usepackage{tablefootnote}

\usepackage[perpage]{footmisc}

\newcommand{\rev}[1]{{\color{black}#1}}

\numberwithin{equation}{section}

\input{newc}
\usepackage{xr}

\title{Surrogate-based global sensitivity analysis with statistical guarantees via floodgate}
\author{Massimo Aufiero and Lucas Janson}
\affil{Department of Statistics, Harvard University}
\date{}

\begin{document}

\maketitle

\begin{abstract}
\input{abstract}
\end{abstract}

\input{introduction}
\input{methods}

\input{simulations}

\input{application}
\input{conclusion}
\input{acknowledgements}
\bibliography{references} 
\bibliographystyle{plainnat}

\newpage

\appendix
\section{Appendix}
\input{appendix/lower_bound_numerator}
\input{appendix/bounds}

\input{appendix/estimator}

\input{appendix/validity}
\input{appendix/width}

\input{appendix/hymod}
\input{appendix/CBMZ}
\end{document}

%% file: newc.tex
\newtheorem{theorem}{Theorem}[section]

\newtheorem{lemma}[theorem]{Lemma}
\theoremstyle{definition}
\newtheorem{definition}[theorem]{Definition}

\newcommand{\E}[1]{\mathbb{E}\left[#1\right]}
\newcommand{\var}[1]{\mathrm{Var}\left(#1\right)}
\newcommand{\cov}[2]{\mathrm{Cov}\left(#1,#2\right)}
\newcommand{\pr}[1]{\mathbb{P}\left(#1\right)}
\def\sumn{\sum\limits_{i=1}^n}
\def\sumK{\sum\limits_{k=1}^K}
\newcommand{\Xcomp}{Z}
\newcommand{\eqas}{\overset{a.s.}{=}}
\def\independenT#1#2{\mathrel{\rlap{$#1#2$}\mkern2mu{#1#2}}}
\newcommand{\indp}{\protect\mathpalette{\protect\independenT}{\perp}}
\newcommand{\fstar}{f^*}
\newcommand{\convp}{\xrightarrow{p}}

\newcommand{\convd}{\xrightarrow{\mathcal{L}}}
\newcommand{\ellhat}{\hat{\ell}_n(f)}
\newcommand{\uhat}{\hat{u}_n(f)}
\newcommand{\Shat}{\hat{S}}
\newcommand{\Lower}{L_n}
\newcommand{\Upper}{U_n}
\newcommand{\R}{\mathbb{R}}
\newcommand{\Rd}{\mathbb{R}^d}
\newcommand{\fZ}{f_z}
\newcommand{\fZfull}{\E{f(X,Z)|\Xcomp}}
\newcommand{\fZstar}{\fstar_z}
\newcommand{\hj}{h}
\newcommand{\hstar}{h^*}
\newcommand{\fZfullstar}{\E{\fstar(X,Z)|\Xcomp}}
\newcommand{\defeq}{:=}
\newcommand{\Xnew}{\tilde{X}_i^{(k)}}
\newcommand{\Xtilde}{\tilde{X}}
\newcommand{\Xtildek}{\Xtilde^{(k)}}
\newcommand{\Xtildej}{\Xtilde_i}
\newcommand{\Xcompj}{\Xcomp_i}

\newcommand{\Xj}{(X_i,Z_i)}
\newcommand{\Xij}{X_i}
\newcommand{\Pcond}{P_{\Xij\mid\Xcompj}}

\newcommand{\Y}{\fstar(X,Z)}
\newcommand{\Ysquared}{{\fstar}^2(X,Z)}
\newcommand{\Yj}{\fstar\Xj}
\newcommand{\Vj}{V_i}
\newcommand{\W}{{M^z}}
\newcommand{\Wj}{M_i^z}
\newcommand{\Zj}{M_i}
\newcommand{\Fj}{{F^z}}
\newcommand{\Fij}{F_i^z}
\newcommand{\Mjbar}{\bar{M}^z}

\newcommand{\ozone}{O3}

%% file: abstract.tex
Computational models are utilized in many scientific domains to simulate complex systems. Sensitivity analysis is an important practice to aid our understanding of the mechanics of these models and the processes they describe, but performing a sufficient number of model evaluations to obtain accurate sensitivity estimates can often be prohibitively expensive. In order to reduce the computational burden, a common solution is to use a surrogate model that approximates the original model reasonably well but at a fraction of the cost. However, in exchange for the computational benefits of surrogate-based sensitivity analysis, this approach comes with the price of a loss in accuracy arising from the difference between the surrogate and the original model. To address this issue, we adapt the floodgate method of \cite{Zhang2020} to provide valid surrogate-based confidence intervals rather than a point estimate, allowing for the benefit of the computational speed-up of using a surrogate that is especially pronounced for high-dimensional models, while still retaining rigorous and accurate bounds on the global sensitivity with respect to the \emph{original} (non-surrogate) model. Our confidence interval is asymptotically valid with almost no conditions on the computational model or the surrogate. Additionally, the accuracy (width) of our confidence interval shrinks as the surrogate's accuracy increases, so when an accurate surrogate is used, the confidence interval we report will correspondingly be quite narrow, instilling appropriately high confidence in its estimate. We demonstrate the properties of our method through numerical simulations on the small Hymod hydrological model, and also apply it to the more complex CBM-Z meteorological model with a recent neural-network-based surrogate.

%% file: introduction.tex
\section{Introduction}
\label{introduction}

\subsection{Problem Statement}
\label{sec: problem}
The use of computational models to describe and simulate complex phenomena is ubiquitous in many scientific fields, and such models play a critical role in both understanding natural processes and informing high-stakes decisions across many domains. Complex systems are first represented with a mathematical model\textemdash usually implicitly using coupled differential equations\textemdash and are then implemented in computer code to exactly or approximately solve these equations. While these models are very useful for simulating processes under various conditions that may be  impossible or infeasibly expensive to observe in reality, the models themselves can often be incredibly computationally expensive, with a single run taking very powerful computers on the order of hours or even days to evaluate, prohibiting the generation of large amounts of data \citep{Razavi2012Numerical}.

Computational models typically rely on a large number of parameters, some of which may be observed data or known physical constants, while many are unobservable quantities for which only their distributions or ranges are known. The large number of uncertain inputs and the complexity of the model results in high variability in the outputs \citep{Do2020,Xu2008}. Thus, the question of how variation in each input factor affects the model outputs is of great interest for a number of reasons. It can be informative both to developers of the model determining how best to improve its structure or reduce uncertainty, as well as to researchers and decision-makers studying the model as a proxy for reality.

Sensitivity analysis is a set of mathematical methods for analyzing how the inputs of a system influence its output(s) by quantitatively measuring the attribution of output variability to uncertainty in the inputs \citep{Pianosi2015}. Sensitivity analysis is a critical component of understanding the intricacies of a model's behavior. However, due to the complexity of the models, it is in most cases impossible to derive an analytical description of input-output relationships, especially since the model itself is almost  never given in closed-form. Thus, most methods for estimating input sensitivities rely on querying the model many times while varying the input values, often with very specific sampling schemes (see Section~\ref{sec: related}). However, obtaining a large enough number of samples from the model to reach a desired level of accuracy is often prohibitively expensive, especially when it needs to be done repeatedly for a large number of inputs. Much of the research on sensitivity analysis is therefore aimed at reducing the total number of model evaluations needed to get accurate estimates.

One popular solution is to use a surrogate model (also referred to as an `emulator' or `metamodel') to approximate the original model at a much lower computational cost. Surrogates can be either data-driven machine learning models or lower-fidelity simulation models that reduce the resolution or number of components of the original. While a computational model's surrogate can
be orders of magnitude faster to run \citep{Kelp2018}, its outputs will in general not be exactly the same as those of the original model when both are given the same inputs.
Therefore, simply applying sensitivity analysis to the surrogate model sacrifices many of the desirable statistical properties, such as consistency and asymptotic normality, with respect to the sensitivity in the original (non-surrogate) model. This means that even though we can generate a much larger number of samples from the surrogate for sensitivity analysis, it is unclear how well the resulting estimates will correspond to the actual sensitivity value of interest.

\subsection{Contribution}
\label{sec: contribution}
We present a novel method for conducting sensitivity analysis that extends \cite{Zhang2020}'s floodgate method for inferring variable importance in high-dimensional regression. Our extension of floodgate (hereafter referred to simply as ``floodgate'') leverages surrogate models for computational efficiency, yet retains statistical guarantees on its estimation with respect to the sensitivity in the \emph{original} model.
In particular, floodgate outputs upper- and lower-bounds for the (original) model sensitivity with provably high asymptotic coverage no matter how accurate the surrogate is. The widths of these intervals directly improve with the accuracy of the surrogate, so that floodgate ensures appropriate uncertainty quantification always, while still providing high precision given a sufficiently high-fidelity surrogate.

Floodgate offers a significant computational advantage compared to non-surrogate-based techniques, 
by a factor of up to the dimension of the input space given a sufficiently fast and accurate surrogate, making it especially advantageous for high-dimensional models.
In addition, given a dataset of sampled inputs and their model evaluations, floodgate can be applied with \emph{no additional} evaluations of the original model (i.e., one only needs to evaluate the surrogate on new inputs), which is not the case for many standard sensitivity analysis techniques since they require very specific pairs of samples. Furthermore, it accounts for surrogate inaccuracy to rigorously quantify the uncertainty of estimates using intervals that are much narrower and require fewer assumptions than existing theoretical bounds on surrogate-based estimation error. Floodgate is applicable to any computational model, any surrogate model, and nearly any input distribution and sampling scheme.

\subsection{Notation}
In this paper, we assume that the computational model $\fstar:\Rd\to\R$ is a deterministic function\textemdash which is true for most computational models\textemdash though even randomized models can be subsumed into our framework by simply including the exogenous randomness (i.e., the random seed) in $\fstar$ as an additional input. Note that we use $\fstar$ to denote the original computational model and $f$ to denote the corresponding surrogate, though in the sensitivity analysis literature the latter is commonly used for computational models. 

We will present floodgate as being applied to a single input of interest, denoted $X$, at a time and we denote the remaining $d-1$ of $\fstar$'s inputs collectively as $Z$. Of course this does not mean floodgate cannot or should not be used for sensitivity analysis of many different inputs, and indeed the more inputs it is applied to, the larger its computational advantage; see Sections \ref{sec: computational_savings}, \ref{sec: simulations}, and \ref{sec: application}.
We use $\Xj$ to denote an individual sample from the input distribution, and we denote the distribution of $\Xij$ conditional on $\Xcompj$ as $\Pcond$. Finally, we use $z_\alpha$ to denote the $\alpha$th quantile of the standard normal distribution.

\subsection{Background and Related Work}
\label{sec: related}
Variance-based global sensitivity analysis methods seek to quantitatively attribute output variance to uncertainty in each input or group of inputs. The estimand that we consider in this paper is the total-order sensitivity index \citep{Homma1996}. This quantity measures the proportion of the total variance of the model output that results from variation in the input $X$, through the sum of all direct effects and interactions with other inputs. A formal definition of this quantity is provided below.  

\begin{definition}{(Total-order sensitivity index)}
\label{def: total_order}
For a computational model $\fstar$, the total-order sensitivity index for input $X$ is defined as
\begin{equation}
    \label{eqn: Si}  
    S \defeq \frac{\E{\var{\Y|\Xcomp}}}{\var{\Y}},
\end{equation}
whenever the appropriate moments exist, with the convention that $0/0=0$.
\end{definition}

Several Monte Carlo (MC) estimators for $S$ have been proposed, including by \cite{Jansen1999}, \cite{Homma1996}, and \cite{Sobol2007}. Most of them utilize the Sobol' pick-freeze (SPF) scheme \citep{Sobol1993,Sobol2001,Janon2013}. In MC estimation, SPF estimators use a set of $n$ i.i.d. pairs of points where within each pair, the values of $\Xcomp$ are the same, but $X$ is sampled (conditionally) independently for both points. To formalize this, assume that for a set of i.i.d. samples $\{(\Xij,\Xcompj)\}_{i=1}^n$ we can sample $\Xtildej\sim \Pcond$ such that $\Xtildej \indp \Xij \mid \Xcompj$ for each $i$. An example SPF estimator for $S$ introduced by \cite{Jansen1999} is
\begin{equation}
	\label{eqn: Sihat}
	\Shat \defeq \frac{\frac{1}{2n}\sumn\big(\Yj - \fstar(\Xtildej,\Xcompj)\big)^2}{\frac{1}{n-1}\sumn \left(\Yj-\frac{1}{n}\sumn\Yj\right)^2}.
\end{equation}
$\Shat$ is consistent for $S$ and was proven by \cite{Sobol2001} and \cite{Saltelli2010} to have lower variance than other similar SPF estimators, so we only focus on this particular $\Shat$ as a comparison to floodgate in Sections~\ref{sec: simulations} and \ref{sec: application}. However, for any of these SPF estimators, computation is very expensive, especially in high dimensions, since computing the full set of $n$-sample sensitivity index estimates for all $d$ inputs requires $n(d+1)$ total evaluations of $\fstar$. Even if one had access to a dataset of i.i.d. input samples with their corresponding model evaluations, $\Shat$ could not be computed without $nd$ additional evaluations of $\fstar$ to create $d$ sets of $n$ paired points as described above. \cite{Sheikholeslami2020} and \cite{Plischke} have developed sensitivity analysis methods applicable to any given data that target different estimands, but to our knowledge there are no such methods for estimating $S$.

As mentioned in Section~\ref{sec: problem}, surrogate modeling techniques are often employed to sidestep this computational obstacle, though they of course sacrifice some accuracy in the obtained estimates. Many surrogates are dynamic models that fully simulate the original model by predicting all of the model's outputs over arbitrary time scales (e.g., \cite{Castelletti2012,Kelp2020}). This includes reduced or simplified versions of the original model (e.g., in \cite{Arandia2019}) or even just running the same model at a lower spatio-temporal resolution. In some cases, a dynamic surrogate is built by using a machine-learning model to replace only a single particularly expensive subcomponent of the original model rather than trying to replace the entire model (e.g., in \cite{Mills2019}) or by using a machine-learning model to map from the outputs of a lower-fidelity simulation to those of the higher-fidelity model (e.g., in \cite{Zhang2019_Nbody}). However, since in sensitivity analysis we are typically only concerned with a particular scalar transformation of the computational model's outputs, data-driven response surface surrogates that learn a direct mapping from the inputs to the output of interest (for some fixed time scale and forcing data, if applicable) can also be used (see \cite{Razavi2012Review} for a review). For any of these types of surrogates, the na\"{i}ve method of estimating $S$ using a surrogate $f$ is to simply replace $\fstar$ with $f$ in the computation of some estimator for $S$. In the case of the SPF estimator $\Shat$ from \eqref{eqn: Sihat}, the surrogate-based estimator is
\begin{equation}
	\label{eqn: Sihatf}
	\Shat^f \defeq \frac{\frac{1}{2n}\sumn\big(f\Xj - f(\Xtildej,\Xcompj)\big)^2}{\frac{1}{n-1}\sumn \left(f\Xj-\frac{1}{n}\sumn f\Xj\right)^2}.
\end{equation} 

Another common use of surrogate models for estimating sensitivity indices that is somewhat less similar to our approach is to fit a special type of response surface surrogate from which estimates of the sensitivity indices can be directly computed from the model coefficients rather than by MC estimation. These methods rely on the fact that $\fstar$ can be exactly represented by an expansion onto some infinite basis, and good approximations of $\fstar$ can often be obtained by using only a finite number of terms in the basis for $f$. The coefficients of the terms in this reduced basis are then used to compute the truncated-sum estimates $\Shat^f$. Two of the most popular such methods are the Fourier amplitude sensitivity test (FAST) \citep{Cukier1973,Saltelli2010} and polynomial chaos expansion (PCE) \citep{Kalinina2020,Cheng2018}, which use Fourier and polynomial bases, as the names suggest.
\cite{Tsokanas2021} use PCE, kriging, and polynomial chaos kriging surrogates for estimating sensitivity indices for a virtual hybrid model representing a prototype motorcycle. \cite{Stephens2011} compare radial basis functions, neural networks, and least squares support vector machines for surrogate-based sensitivity analysis of a computational fluid dynamics model. \cite{Gratiet2017} demonstrate the use of Gaussian processes to obtain confidence intervals on sensitivity indices for a truss structure engineering model, as well as PCE for point estimates.

Both approaches to surrogate-based sensitivity analysis face the same issue: the estimates that use $f$ in place of $\fstar$ do not truly estimate $S$, but rather the quantity
$S^f := \E{\var{f\Xj\mid\Xcomp}}/\var{f\Xj},$
where the difference $S^f - S$ is unknown, making it hard to rigorously quantify the error of the estimates $\Shat^f$ with respect to $S$. This is because there is now a surrogate error (i.e., $S^f-S$) in addition to the sampling error ($\Shat^f-S^f$) in the estimates. \cite{Janon2012} present a method that uses the MC estimates $\Shat^f$ to obtain upper and lower bounds of $\Shat$, but it requires a computationally intensive optimization procedure repeated over bootstrap samples as well as knowledge of the pointwise error bound on $|f(X,Z) - \Y|$, which is only computable for a few specific types of surrogates and computational models. \cite{Janon2013} establish conditions on the rate of convergence of a sequence of surrogates $f_n$ to $\fstar$ in order for $\Shat^{f_n}$ to be consistent for $S$ and asymptotically normal in the double limit as $n\to\infty$ and $f_n\to\fstar$, where the surrogate $f_n$ depends on the sample size (e.g., if some fixed proportion of the data is used for training the surrogate and the rest for computing $\Shat^{f_n}$). However, these conditions require convergence rates of the model error to zero that are only satisfied by a limited class of computational model--surrogate pairs, and thus the results are not applicable in general. Most recently, \cite{Panin2021} provides a bound on the surrogate error $|S^f - S|$ that depends on the mean squared error (MSE) of $f$ and is estimable from data. We compare floodgate to their bounds both theoretically (Section~\ref{sec: compare_methods}) and empirically (Section~\ref{sec: compare_application}) by extending their methodology to obtain confidence intervals as well, demonstrating that floodgate produces intervals that are consistently and substantially narrower than theirs.

As mentioned in Section~\ref{sec: contribution}, our method is an extension of the original floodgate method for high-dimensional inference from \cite{Zhang2020}. Their algorithm outputs an asymptotically valid lower confidence bound for the numerator of $S$ in the general regression setting.
Instead of the lower-bound function used in \cite{Zhang2020}, we use the function introduced in \citep{Zhang2022}, a later work focused on other estimands. \cite{Zhang2020} notes
that the numerator of $S$ can be upper-bounded as well, but that the bound cannot be made tight except in the noiseless setting---an edge case in that paper but the case of primary interest in this paper---so they do not pursue the idea further. 
In contrast to these works, we derive and give full treatment to an upper confidence bound (which is tight, as we work in the noiseless setting) so that we can provide a (two-sided) confidence interval for $S$.
Indeed, for sensitivity analysis, upper confidence bounds are often of even more value than lower confidence bounds, as they allow dimensionality reduction via dropping inputs with low sensitivities. We also present novel results on the asymptotic width of our confidence interval and on the computational speedups floodgate offers compared to non-surrogate-based sensitivity analysis methods, as well as numerical demonstration of floodgate's value for sensitivity analysis and efficient \href{https://github.com/aufieroma12/floodgate}{code}\footnote{Code available at https://github.com/aufieroma12/floodgate} in python to ease the use of floodgate by others.

%% file: methods.tex
\section{Methods}
\label{methods}

\subsection{Bounds for the total-order sensitivity index}
\label{sec: bounds}
As outlined in Section~\ref{sec: related}, there are a number of existing estimators that are consistent and asymptotically normal for $S$ when using samples from the computational model $\fstar$, but when a surrogate model $f$ is substituted for $\fstar$, these properties are no longer guaranteed for $\Shat^f$. Indeed, as detailed in Section~\ref{sec: related}, for a fixed surrogate $f$ (i.e., $f$ does not change with $n$), $\Shat^f$ will converge to a value $S^f$ that is not equal to $S$ in general. Thus, we would like to have a way to leverage a surrogate model $f$ to estimate $S$ \emph{with a computable bound on the error}, ideally as small as possible.

We introduce a bias-aware surrogate-based method for inference on $S$ that allows us to quantify uncertainty in the form of confidence intervals. Floodgate uses surrogate examples to estimate upper and lower bounds of $S$.  In particular, we 
\begin{enumerate}
    \item[(a)] define functions $\ell(f)$ and $u(f)$ such that $\ell(f) \le S \le u(f)$ for any surrogate $f$,
    \item[(b)] construct estimators $\ellhat$, $\uhat$
    that converge to $\ell(f)$ and $u(f)$, respectively,
    \item[(c)] derive a confidence interval $[\Lower,\Upper]$ with provable coverage for $S$.
\end{enumerate}
The intuition behind this process is that if $\Lower$ is a lower confidence bound for $\ell(f)$, then it is by construction also a lower confidence bound for $S$, and similarly for the upper confidence bound $\Upper$.

We use the following upper- and lower-bound functions, whose numerators were originally derived by \cite{Zhang2020} and \cite{Zhang2022}, respectively. For any surrogate $f: \Rd \to \R$, define
\begin{align}
    u(f) &= \frac{\E{\left(\Y - \fZfull\right)^2}}{\var{\Y}} \label{eqn: u_i}\\
    \ell(f) &= u(f) - \frac{\E{(\Y - f(X,Z))^2}}{\var{\Y}}, \label{eqn: ell_i}
\end{align}
again with the convention that 0/0 = 0.\footnote{Note that when $\var{\Y}=0$, the numerator of $u(f)$ is always non-negative (so $u(f)$ may take the value $\infty$) and the numerator of $\ell(f)$ is always non-positive (so $\ell(f)$ may take the value $-\infty$). This is clearly true for $u(f)$ from \eqref{eqn: u_i} and follows for $\ell(f)$ from the proof in Appendix~\ref{apx: lj_numerator}.} The numerator of $u(f)$ is simply the MSE of $\fZ(\Xcomp) \defeq \fZfull$, which is a function of only $\Xcomp$, and the numerator of $\ell(f)$ is the difference between this quantity and the MSE of $f$ itself. Lemma \ref{lem: bounds} is a key result for proving the accuracy and validity of floodgate. The proof of this lemma is provided in Appendix~\ref{prf: bounds}.
\begin{lemma}
\label{lem: bounds}
\textit{For any $\fstar$ and $f$ such that $\ell(f)$ and $u(f)$ exist,}
\begin{equation}
    \ell(f) \le \ell(\fstar) = S = u(\fstar) \le u(f).
\end{equation}
\end{lemma}

\subsection{Estimators and confidence intervals} 
Given samples $\{\Xj\}_{i=1}^n$, we can construct simple MC estimators $\ellhat$ and $\uhat$. We start by defining the quantities
\begin{align}
    \text{MSE}(f) & = \E{(\Y - f(X,Z))^2}\\
    \text{MSE}(\fZ) &= \E{\left(\Y - \fZ(\Xcomp)\right)^2}.
\end{align}
Thus, we can express $u(f) = \frac{\text{MSE}(\fZ)}{\var{\Y}}$ and $\ell(f) = \frac{\text{MSE}(\fZ) - \text{MSE}(f)}{\var{\Y}}$. If the samples $\Xj$ are i.i.d., then we easily obtain unbiased MC estimators for each term in the numerators and denominators by simply replacing the outer expectations with sample means. For MSE$(f)$, this is straightforward. We can compute the sample mean of 
\begin{equation}
\label{eqn: Zij}
    \Zj \defeq \left(\Yj - f\Xj\right)^2,
\end{equation}
for $i\in \{1,\dots,n\}$ and use this as our estimator. Note that $\Zj$ is a (random) function of $f$, but we drop the dependence on $f$ in the notation for simplicity. $\Zj$ is trivially unbiased for MSE$(f)$. However, unbiased estimation of MSE$(\fZ)$ and $\var{\Y}$ is somewhat less straightforward as both are expected squared errors with respect to expectations which are themselves generally intractable to compute analytically.
For $\var{\Y}$, we can use the standard unbiased variance estimator, which is the sample mean of 
\begin{equation}
\label{eqn: Vj}
    \Vj \defeq \frac{n}{n-1} \left(\Yj - \frac{1}{n}\sum_{i'=1}^n\fstar(X_{i'},Z_{i'})\right)^2.
\end{equation}
For MSE$(\fZ)$, the challenge is dealing with the analytically intractable $\fZ(\Xcomp)=\fZfull$. Assuming that we can sample $K \ge 1$ copies $\Xnew$ of $\Xij$ from the conditional distribution $P_{\Xij|\Xcompj}$, where each $\Xnew$ is conditionally independent of $\Xij$, then for each $i\in \{1,\dots,n\}$ we can estimate $\fZ(\Xcompj)$ with 
$$\frac{1}{K}\sum_{k=1}^K f(\Xnew, \Xcompj),$$
which we will use in our estimator for MSE$(\fZ)$. In particular, the quantity 
\begin{equation}
\label{eqn: Wij}
    \Wj \defeq \left(\Yj - \frac{1}{K}\sum_{k=1}^K f(\Xnew, \Xcompj)\right)^2 - \frac{1}{K+1}\left(f\Xj - \frac{1}{K}\sum_{k=1}^K f(\Xnew, \Xcompj)\right)^2
\end{equation}
is unbiased for MSE$(\fZ)$. This is formalized in Lemma \ref{lem: unbiased}, the proof of which is provided in Appendix~\ref{prf: unbiased}. Again, we drop the dependence of $\Wj$ on $f$ in the notation for simplicity.

\begin{lemma}
\label{lem: unbiased}
\textit{For $K\ge 1$, given a set of i.i.d. original samples $\{\Xj\}_{i=1}^n$ and modified samples $\tilde{X}_i^{(1)},\dots, \tilde{X}_i^{(K)} \sim \Pcond$ such that $\Xnew \indp \Xij | \Xcompj$ for each $i,k$, then for any $\fstar, f: \Rd\to\R$, the quantity $\Wj$ defined in \eqref{eqn: Wij} satisfies:}
\begin{equation*}
    \E{\Wj} = \textnormal{MSE}(\fZ).
\end{equation*}
\end{lemma}

Since $\{(\Wj , \Zj, \Vj)\}_{i=1}^n$ are i.i.d. and unbiased for $(\text{MSE}(\fZ), \text{MSE}(f), \var{\Y})$, their sample means $\Mjbar$, $\bar{M}$, and $\bar{V}$ are asymptotically normal estimators for their respective estimands. We can then combine these to get estimators for $\ell(f)$ and $u(f)$ that are also asymptotically normal by the delta method. In particular, the estimators are 
\begin{align}
    \ellhat &= \frac{\Mjbar - \bar{M}}{\bar{V}} \label{eqn: ellhat}\\
    \uhat &= \frac{\Mjbar}{\bar{V}}. \label{eqn: uhat}
\end{align}

Due to the asymptotic normality of these estimators, it is straightforward to obtain asymptotically valid upper and lower confidence bounds, as described in Algorithm \ref{alg: conf}. 

\begin{algorithm}[H]
\caption{Floodgate for surrogate-based sensitivity analysis}
\label{alg: conf}
\begin{algorithmic}
    \REQUIRE Samples $\{\Xj\}_{i=1}^n$, surrogate $f: \Rd\to\R$, $K\in\mathbb{N}$, and confidence level $\alpha\in (0,1)$.\\
    For each $i\in \{1,\dots,n\}$, compute $\Wj$, $\Zj$, $\Vj$ according to \eqref{eqn: Wij}, \eqref{eqn: Zij}, and \eqref{eqn: Vj}, their sample means $(\Mjbar, \bar{M}, \bar{V})$, and their $3\times 3$ sample covariance matrix $\hat{\Sigma}$. \\
    \rev{If $\bar{V}=0$, set $\Lower=0$ and $\Upper=1$. Else, compute
    \begin{align*}
    		s_u^2 &= \frac{1}{\bar{V}^2}\left(\hat{\Sigma}_{11} - 2\frac{\Mjbar}{\bar{V}}\hat{\Sigma}_{13} + \Big(\frac{\Mjbar}{\bar{V}}\Big)^2\hat{\Sigma}_{33}\right)\\
    		\text{and } \; s_\ell^2 &= \frac{1}{\bar{V}^2}\left(\hat{\Sigma}_{11} + \hat{\Sigma}_{22} + \left(\frac{\Mjbar - \bar{M}}{\bar{V}}\right)^2\hat{\Sigma}_{33} - 2\hat{\Sigma}_{12} + 2\frac{\Mjbar-\bar{M}}{\bar{V}}\left(\hat{\Sigma}_{23} - \hat{\Sigma}_{13}\right)\right), \\
    		\text{and set } \; \Lower &= \max\left\{0, \frac{\Mjbar-\bar{M}}{\bar{V}} - \frac{z_{\alpha/2} s_\ell}{\sqrt{n}}\right\} \quad \text{and} \quad \Upper = \min\left\{1, \frac{\Mjbar}{\bar{V}} + \frac{z_{\alpha/2} s_u}{\sqrt{n}}\right\}.
    \end{align*}
    \ENSURE Confidence interval $[\Lower,\Upper]$.}

\end{algorithmic}
\end{algorithm}

Theorem \ref{thm: validity} establishes the asymptotic coverage of the interval $[\Lower, \Upper]$, and the proof is given in Appendix~\ref{prf: validity}. It requires only very mild moment assumptions that are standard for the central limit theorem (CLT). We need $\fstar(X,Z)$ and $f(X,Z)$ to have finite fourth moments rather than the standard second moments because of the squared terms in the estimators and estimands.

\begin{theorem}
\label{thm: validity}
\textnormal{(Asymptotic coverage).} For i.i.d. samples $\{\Xj\}_{i=1}^n$, any computational model $\fstar$ and surrogate $f:\Rd\to\R$, and any $\alpha\in (0,1)$, if $\E{{\fstar}^4(X,Z)}, \E{f^4(X,Z)} < \infty$, then the bounds $\Lower$ and  $\Upper$ output by Algorithm~\ref{alg: conf} satisfy 
$$\liminf_{n\rightarrow\infty}\pr{\Lower \le S \le \Upper} \ge 1 - \alpha.$$
\end{theorem}

Note that the validity of the confidence interval derived from $f$ does not depend on $f$ itself. We do not require any conditions on the quality of $f$\textemdash in fact, a less accurate $f$ would tend to have higher coverage, since $\ell(f)$ and $u(f)$ would span $S$ by a wider margin, whereas the confidence bounds derived from a na\"{i}ve surrogate-based estimator such as $\Shat^f$ in \eqref{eqn: Sihatf} are not guaranteed to cover the true value at all when $f$ differs from $\fstar$. 

Of course, in practice we hope that $f$ \emph{is} an accurate surrogate for $\fstar$. Since there are very few assumptions on the properties of $f$, we can leverage state-of-the-art machine learning algorithms and arbitrary domain knowledge to construct a surrogate that is as accurate as possible. For example, $f$ can be any of the various types of machine learning, physically-based, or hybrid surrogate models described in Section~\ref{sec: related}, with the only restriction being that it is independent of the data used for constructing the floodgate bounds (e.g., if $f$ is fitted via machine learning, its training data should be independent of the data used to compute $\Lower$ and $\Upper$). Theorem \ref{thm: width} establishes that the width of the confidence interval $[\Lower, \Upper]$ converges to a value depending on the accuracy of $f$, and it does so at an $O_p(n^{-1/2})$ rate. Thus, when $f$ is very close to $\fstar$, we can asymptotically achieve very tight bounds while still guaranteeing coverage. The proof of Theorem \ref{thm: width} is provided in Appendix~\ref{prf: width}.
\begin{theorem}
\label{thm: width}
\textnormal{(Width of confidence intervals).} 
Under the same assumptions as in Theorem~{\ref{thm: validity}} and the additional assumption that $\var{\Y}>0$, the bounds $\Lower$ and  $\Upper$ output by Algorithm~\ref{alg: conf} satisfy
$$\Upper - \Lower \rev{\le} \frac{\textnormal{MSE}(f)}{\var{\Y}} + O_p\left(n^{-1/2}\right).$$
\end{theorem}

\subsection{Computational savings}
\label{sec: computational_savings}
We consider here the computational expense of floodgate when applied to \emph{every} input of $\fstar$. Thus, for the remainder of this section (and in Sections~\ref{sec: simulations} and \ref{sec: application}), we will explicitly label the total-order sensitivity index for the $j$th input using $S_j$ (i.e., $S_j$ corresponds to labelling the $j$th input as $X$).

For an individual sensitivity index, Algorithm~\ref{alg: conf} requires a set of only $n$ points $\Xj$ evaluated by $\fstar$, plus an additional $K$ samples $(\Xnew,Z_i)$ evaluated by $f$ for each $i\in \{1,\dots,n\}$, for a total of $nK$ evaluations of $f$. While these $nK$ evaluations of $f$ on the resampled inputs are distinct for each of the $d$ sensitivity indices, the \emph{same} set of $n$ original points evaluated by $\fstar$ are used every time.
Since we generally assume that the surrogate is much less expensive to evaluate than $\fstar$, we expect the cost of the $ndK$ total evaluations of $f$ necessary for inference on the full set of $\{S_j\}_{j=1}^d$ is negligible compared to the $n$ total evaluations of $\fstar$ necessary. If we need to train the surrogate ourselves first, this will require an additional $n_{train}$ evaluations of $\fstar$.

For comparison, $n$-sample estimation of all $d$ sensitivity indices by $\Shat$ requires $n(d+1)$ total evaluations of $\fstar$. Note that when $K=1$ and $f=\fstar$, $\ellhat=\uhat=\Shat$, so in the ideal case where we have a surrogate that is perfectly accurate and takes zero computation time, the output of floodgate with $K=1$ is identical to estimation/inference using $\Shat$ but with computation time lower by a factor of $1/(d+1)$ (and floodgate will only be more accurate with larger $K$). In general (continuing to assume $K=1$ for simplicity), if we let $c$ denote how many times higher $\fstar$'s computational expense is than $f$'s, the same speedup approximately holds as long as 
$c \gg d$
and as long as MSE$(f)/\var{\Y}\ll n^{-1/2}$. If one or both of these conditions do not hold, the computational advantage of floodgate will be less than a factor of $1/(d+1)$, though the speedup may still be significant. Floodgate's advantage disappears roughly when 
MSE$(f)/\var{\Y} \gtrsim \left(n(\frac{1}{c}+\frac{1}{d})\right)^{-1/2}$, which makes intuitive sense since it means the surrogate is either not particularly fast or not particularly accurate (or both), and naturally we should not expect to be able to leverage such a low-quality surrogate to beneficial effect.

Another advantage of floodgate compared to most non-surrogate methods is that it can be applied to almost any pre-existing dataset\textemdash which may have been collected for a  purpose unrelated to sensitivity analysis or obtained from another source\textemdash without the need for additional evaluations of $\fstar$. While we present our results for the case of i.i.d. data, they generalize to any sampling scheme that is compatible with asymptotically normal estimation of expectations of functions of the samples. For example, CLTs exist for Latin hypercube sampling \citep{Owen1992} and general randomized quasi-Monte Carlo sampling techniques \citep{Owen2019}, which are frequently used in sensitivity analysis because they achieve faster rates of convergence than standard MC estimation. Indeed, we demonstrate the application of floodgate to a dataset of non-i.i.d. samples in Section~\ref{sec: application}.

\subsection{Comparison to existing bounds on error of $\Shat^f$}
\label{sec: compare_methods}
As mentioned in Section~\ref{sec: related}, \cite{Panin2021} proved bounds on the surrogate error term $|S^f-S|$ and provide a method for estimating those bounds. In particular, Corollary~1 of their paper establishes the bound
\begin{equation}
	\label{eqn: panin}
	|S^f - S| \le \min\left\{1, \mathcal{E} + 2\sqrt{S}, \mathcal{E} + 2\sqrt{1-S}\right\}\mathcal{E},
\end{equation}
where $\mathcal{E} = \sqrt{\text{MSE}(f)/\var{\Y}}$. Therefore, adding and subtracting the bound on the right-hand side of \eqref{eqn: panin} to the surrogate sensitivity $S^f$ yields an interval guaranteed to contain the true sensitivity $S$. Note that if $S$ is not exactly 0 or 1, then as $f$ converges to $\fstar$, the width of this interval is $O(\mathcal{E})$.

The interval $[\ell(f), u(f)]$ that we provide in Section~\ref{sec: bounds} is also guaranteed to contain the true sensitivity $S$, but its width is only $O(\mathcal{E}^2)$, so it shrinks at a much faster rate as the quality of the surrogate improves (i.e., MSE$(f) \to 0$). In fact, since its width is \emph{exactly} $\mathcal{E}^2$ (see \eqref{eqn: ell_i}),
by comparing $\mathcal{E}^2$ to each of the three terms in the bound in \eqref{eqn: panin}, it is immediate that our interval $[\ell(f), u(f)]$ is strictly narrower than that derived from \eqref{eqn: panin} whenever $S \notin \{0,1\}$ and when $\mathcal{E} < 1$, the latter of which we would expect to be true for any reasonable surrogate.

\cite{Panin2021} also propose a natural plug-in estimator for the bound in \eqref{eqn: panin} in Section~3.1.2., which we note can be extended to obtain confidence intervals as well via a similar CLT argument to what we use. While computing $\Shat^f$ requires no evaluations of $\fstar$ (or $n_{train}$ evaluations if $f$ must be trained from scratch), one still needs $n>1$ samples from $\fstar$ in order to estimate $\mathcal{E}$. Thus, computing confidence intervals for a set of $d$ sensitivity indices requires $n$ evaluations of $\fstar$ (and $nd$ evaluations of $f$), similar to floodgate.

We perform empirical comparisons of our bound to \cite{Panin2021}'s in Sections \ref{sec: compare_application} and \ref{sec: application}.

%% file: simulations.tex
\section{Simulations}
\label{sec: simulations}

\subsection{Overview of computational model}
\label{sec: hymod_overview}
We conducted numerical experiments using the Hymod \citep{Boyle2001, Wagener2001} hydrological model to demonstrate floodgate's guarantees on coverage and the relative widths of its confidence intervals compared to standard SPF estimators. Hymod is relatively simple, low-dimensional, and inexpensive to evaluate (though still slower than most common surrogates). Thus, it would not likely be necessary to use a surrogate for it and it would not be the ideal target of floodgate in practice; however, it was a good candidate for these simulations \textit{because} it is so inexpensive to evaluate, meaning we were able to run many independent trials and obtain precise approximations of the sensitivity indices for reference.
In addition, Hymod has been used frequently within the sensitivity analysis literature as a test case for new methods (e.g., \cite{Herman2013, Razavi2016, Cheng2018, Sheikholeslami2020}).

There are five uncertain parameters in the model governing the mechanics of the system, and we treat these as the model inputs whose sensitivities we study. The names, descriptions, units, and ranges of these inputs are given in Table~\ref{tab: hymod_parameters} in Appendix~\ref{apx: hymod}. We used forcing data and observed outputs obtained from the Leaf Catchment in Mississippi \citep{Pianosi2016} in these experiments. An implementation of Hymod in Python was obtained from the \href{https://www.safetoolbox.info}{SAFE Toolbox} \citep{Pianosi2016}.

\subsection{Simulation setup}
\label{sec: experiments}
The response variable we considered for the Hymod simulations was the Nash--Sutcliffe efficiency criterion (NSE) \citep{Nash1970}. The NSE is a very commonly used metric in hydrology to assess how well model predictions agree with observed data. It is often used in sensitivity analysis studies, since it helps understand what inputs significantly impact model accuracy and thus should be calibrated more carefully, or what inputs have a very small effect and could thus be fixed or dropped for model simplification.

Since we need to be able to draw i.i.d. samples $\Xj$ and conditionally i.i.d. samples $\Xnew$ for each sensitivity index, we must define the joint input distribution. As is standard in other global sensitivity analysis studies of this model, we assumed that the inputs were independent and followed uniform distributions. The ranges we used for each variable's distribution are provided in Table~\ref{tab: hymod_parameters} in Appendix~\ref{apx: hymod}.

We applied floodgate to compute confidence intervals for the full set of $d=5$ sensitivity indices using different computational budgets $N$ ranging from $100$ to $50000$ (to be defined shortly). We compare our intervals to asymptotically normal confidence intervals derived from $\Shat$ \eqref{eqn: Sihat} and $\Shat^f$ \eqref{eqn: Sihatf}, clipping the bounds to be within $[0,1]$. In Section~\ref{sec: compare_application} we also provide a more detailed comparison to the error bounds from \cite{Panin2021}. We define the computational budget as the total number of evaluations of $\fstar$ required in the full process of estimating all $d$ sensitivity indices. \begin{itemize}
    \item For floodgate, the full set of $n=N$ evaluations of $\fstar$ is used for each input. Recall that floodgate uses the same set $\{\fstar\Xj\}_{i=1}^n$ and distinct sets of $nK$ evaluations of $f$ for each input.
    \item For the non-surrogate SPF estimator $\Shat$, only $n=N/(d+1)$ pairs of evaluations of $\fstar$ are used for each input. Recall that computation of each $\Shat$ uses the same set $\{\fstar\Xj\}_{i=1}^n$ and a distinct set $\{\fstar(\Xtildej,\Xcompj)\}_{i=1}^n$ for each input.
\end{itemize}
This relationship between $n$ and $N$ puts floodgate and $\Shat$ on an equal computational footing, assuming that  evaluation of $f$ is negligible compared to $\fstar$. However, since we use pretrained surrogates, it is not possible to do the same for $\Shat^f$ as it uses no evaluations from $\fstar$. We choose to use $n=N$ pairs of evaluations of $f$ for each input. While $n$ could have been chosen to be arbitrarily large for $\Shat^f$, the main conclusion about $\Shat^f$ from these experiments is its lack of coverage, which is only more evident at larger sample sizes.

Note that if one did not have a pretrained surrogate and had to train one from scratch, then for floodgate, some fraction of the $N$ samples (e.g., $N/2$) would have to be reserved for training, while the remaining $n=N/2$ are used for computing the confidence intervals. In this case, the surrogate-based SPF estimator $\Shat^f$ would use all $N$ samples to train and 0 samples from $\fstar$ plus an arbitrary number of samples from $f$ for inference. The relationship between $n$ and $N$ remains unchanged for $\Shat$ since it does not use a surrogate.

To simulate having surrogates of various qualities, we conducted all experiments using kernel ridge regression (KRR) with radial basis function kernel pretrained on different amounts of data. In particular, we train low-quality (MSE$(f)/\var{\Y} \approx 0.07$; dashed lines) and high-quality (MSE$(f)/\var{\Y} \approx 0.01$; solid lines) surrogates. In practice, we would expect the type of surrogates used to incorporate specific domain knowledge and potentially to have been trained offline and made publicly available, rather than be fitted by an out-of-the-box machine learning model as done here. However, Hymod is simple and low-dimensional enough that there are no existing high-fidelity surrogates, so fitting a nonparametric model on a large separate dataset provided a reasonable alternative for obtaining surrogates to test our method.

\subsection{Results}
\begin{figure}
\begin{center}
\includegraphics[width=\textwidth]{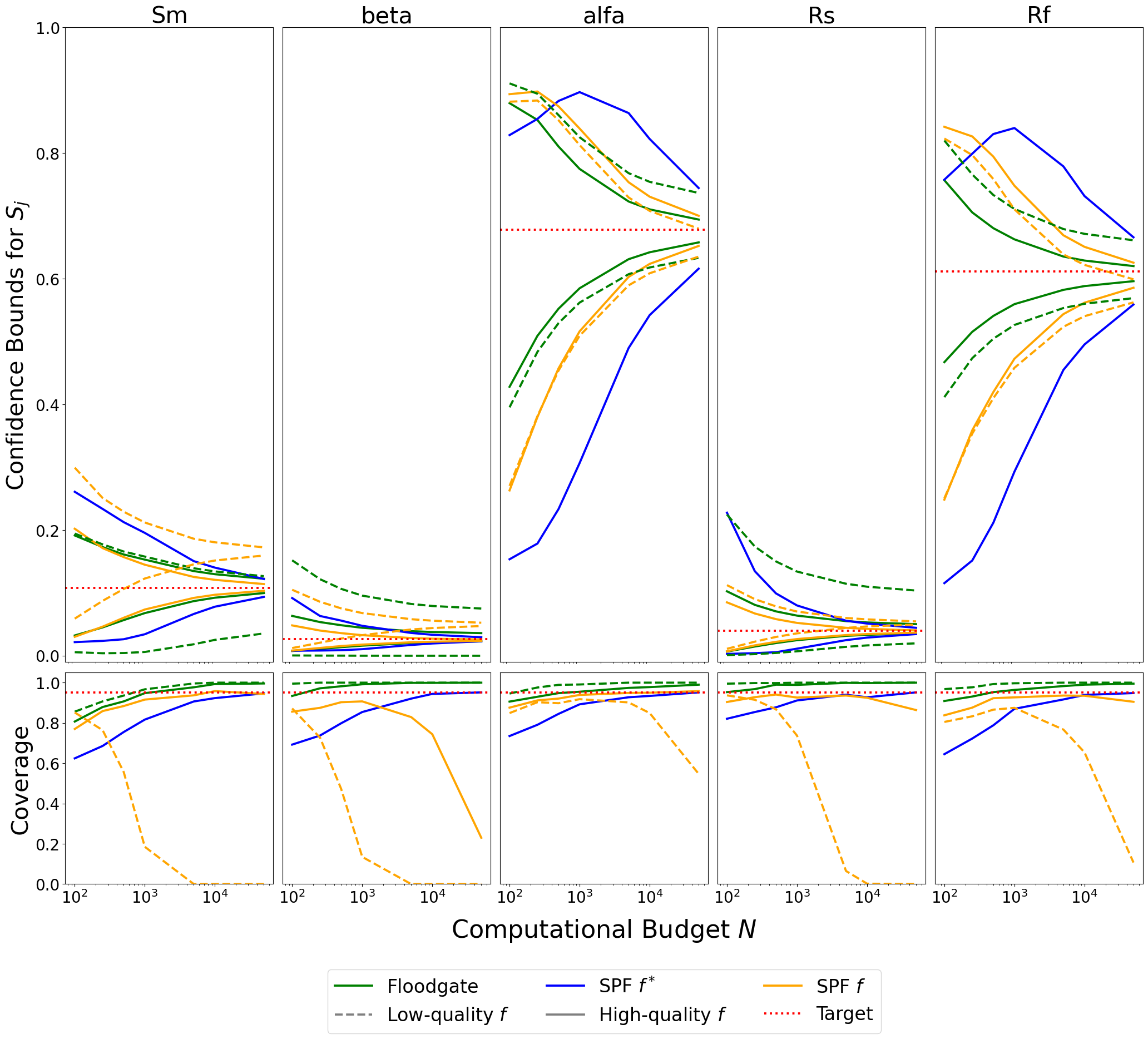}
\end{center}
\caption{95\% confidence intervals and empirical coverage for the Hymod model sensitivity indices using floodgate and the non-surrogate- and surrogate-based SPF estimators with different quality surrogates.}
\label{fig: hymod_pretrained_conf}
\end{figure}

As mentioned in Section~\ref{sec: computational_savings}, we will index $S_j$ (and similarly for its estimators) here to explicitly denote the total-order sensitivity index for the $j$th input, since we apply floodgate and other methods to every input. Figure~\ref{fig: hymod_pretrained_conf} shows the confidence intervals obtained by floodgate and the two other methods, along with their empirical coverage. The curves in the confidence bound plots are the averages over 1000 independent trials. All standard errors are less than 0.008 on the left-hand sides of the plots and less than 0.001 on the right-hand sides. The horizontal dotted red lines in the confidence bound plots represents the ``ground truth'' $S_j$ values, which were estimated using a consistent estimator with $10^8$ samples. These ground truth values were used for computing the coverage. The horizontal dotted red lines in the coverage plots represent the nominal level of 95\%.

These plots demonstrate that floodgate can output narrow (and valid) confidence intervals when using an accurate surrogate. Our high-quality intervals (solid green curves) are almost always tighter than those using the non-surrogate SPF estimator (blue curves) for smaller sample sizes. For example, for the input Rf, our estimated interval is $[0.467, 0.757]$ for 100 model evaluations, whereas the non-surrogate-based method gives the much wider interval $[0.116 0.758]$. For the input Rs, we provide an upper confidence bound of $0.102$ for $N=100$, whereas the non-surrogate-based method's is more than double that ($0.228$). 

In addition, our coverage is almost always valid with only a few small violations at the smallest sample sizes, and it is consistently higher than the other methods', \textit{even} when our intervals are narrower. Some of the reasons contributing to this is that each of our estimators gets to use $d+1 = 6$ times as many samples evaluated by $\fstar$ as $\Shat_j$ due to the computational budget constraints (putting floodgate closer to ``asymptopia") and our bounds are conservative against the bias of $f$. While the surrogate-based SPF estimator (orange curves) also outputs narrow confidence intervals, they fail to account for surrogate inaccuracy and thus have no guarantees on validity, as demonstrated in the coverage plots. Thus, this estimator can output very high confidence for an incorrect estimate. As an example, on the right-hand side of the plot for the input Sm, the low-quality surrogate-based SPF method outputs a very narrow interval around a value roughly \emph{double} the true value\textemdash the coverage here is of course zero.

As mentioned in Section~\ref{sec: hymod_overview}, these results are all for a low-dimensional model where the advantage of floodgate should be particularly \emph{un}pronounced. We were able to achieve much narrower intervals (while maintaining valid coverage) than the non-surrogate-based method with just $6$ times the number of samples here, but for a higher-dimensional model we would expect to gain a larger advantage---we will demonstrate an application of floodgate to a roughly 100-dimensional model in Section~\ref{sec: application}.

\subsection{Comparison to Existing Bounds on the Error of $\Shat^f$}
\label{sec: compare_application}

\begin{figure}
\begin{center}
\includegraphics[width=\textwidth]{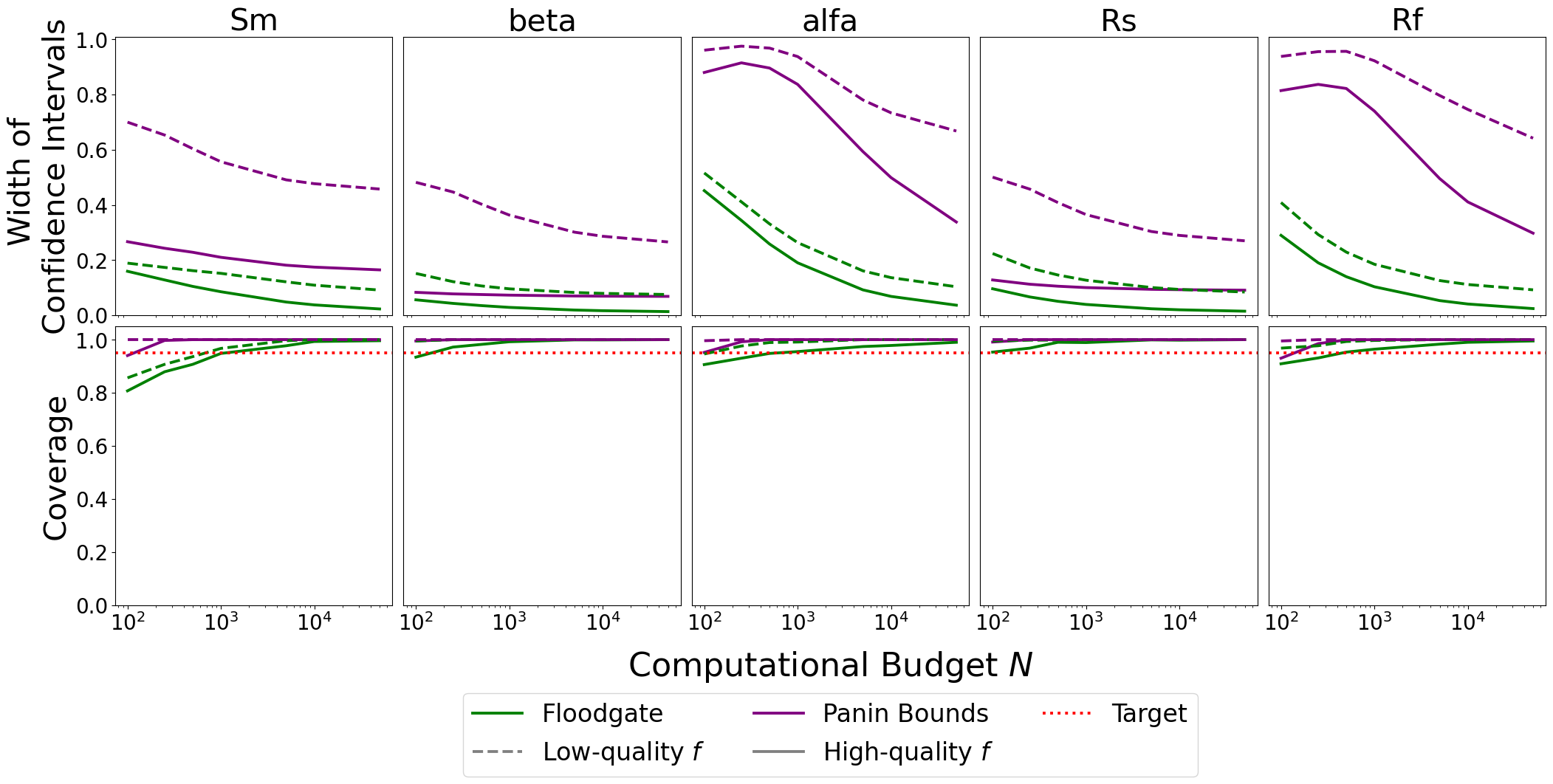}
\end{center}
\caption{Widths of 95\% confidence intervals and empirical coverage for the Hymod model sensitivity indices using floodgate and \cite{Panin2021}'s bounds with different quality surrogates.}
\label{fig: hymod_pretrained_panin}
\end{figure}

We also compare floodgate to the bounds derived by \cite{Panin2021}, discussed in Section~\ref{sec: compare_methods}. In particular, we compare our confidence intervals for each $S_j$ to those computed by adding and subtracting their estimated bound on surrogate error to $\Shat_j^f$ and applying the CLT with the multivariate delta method, clipping the bounds to be within $[0,1]$. Figure~\ref{fig: hymod_pretrained_panin} shows the width of each method's confidence intervals averaged over 1000 independent trials along with their empirical coverage for different values of the computational budget $N$ and using low-quality and high-quality surrogates. All standard errors are less than 0.0083 on the left-hand sides of the plots and less than 0.0004 on the right-hand sides. The dependence on $N$ for floodgate (green curves) is the same as in the previous section. For the confidence intervals based on \cite{Panin2021}'s bounds (purple curves), all $N$ evaluations of $\fstar$ are used to estimate $\mathcal{E}$, and we again use $n=N$ pairs of samples from $f$ for computing $\Shat_j^f$. Thus, both methods use $N$ total evaluations from $\fstar$. We chose to plot the widths of the intervals rather than the intervals themselves on the same set of axes as in Figure~\ref{fig: hymod_pretrained_conf} because both methods give the same guarantees on coverage of $S_j$, and thus the positions of the intervals are less interesting than their relative widths. Indeed, both methods' empirical coverage is valid almost everywhere, reaching practically 100\% on the right-hand side of the plots since both methods' bounds are conservative.

We see that the width of our confidence intervals are consistently substantially smaller than those of \cite{Panin2021} for each sensitivity index and surrogate.
The empirical coverage for both methods is above the nominal level in nearly every setting. Floodgate's coverage does dip below most notably for the input Sm for smaller $N$ values, though even in this case it never falls below 80\%. The intervals derived using \cite{Panin2021}'s bounds tend to have higher coverage than the floodgate interval, which is expected given that  their bounds are looser in the accurate-surrogate regime as shown in Section~\ref{sec: compare_methods}. Coverage for both methods becomes consistently $\ge$95\% as $N$ increases, as they both have the same asymptotic guarantees. 

We thus demonstrate that floodgate achieves rigorous quantification of the uncertainty of surrogate-based estimates using tighter bounds than those provided by \cite{Panin2021}.

%% file: application.tex
\section{Application}
\label{sec: application}

\subsection{Overview of computational model and surrogate}
To demonstrate a more realistic application of floodgate, we used the Carbon Bond Mechanism Z (CBM-Z) meteorological model \citep{Zaveri1999} for simulating tropospheric gas-phase chemistry. It is both higher-dimensional and more computationally expensive than Hymod, and there are existing surrogates that have been built for it. CBM-Z models the evolution of 101 gaseous and aerosol species over a given time period by numerically integrating a system of partial differential equations. In addition to the initial concentrations of each species, the model has four additional meteorological inputs: temperature, pressure, relative humidity, and the cosine of the solar zenith angle. While we use a stand-alone version of CBM-Z for our experiments, it is commonly employed as the gas-phase chemistry module within larger chemistry transport models used for air quality forecasting, and it is typically the most time-consuming component \citep{Wang2019}. 

For our experiments, we used a multitarget regression neural network surrogate from \cite{Kelp2020}. The network consists of an encoder that reduces the input concentrations to a lower-dimensional latent representation, an operator that approximates the integration step in the latent space, and a decoder that maps the integrated system back to the original space. The operator can be applied recurrently to make predictions over arbitrary time scales. For a 24-hour simulation, \cite{Kelp2020} report a speedup by a factor of roughly $3700$ compared to the true CBM-Z model on the same hardware, while maintaining high accuracy for various outputs of interest on an independent test set of randomly sampled inputs.

\subsection{Experimental setup}
We consider the predicted concentration of ozone (\ozone) over a 2-hour interval as the response variable of interest. Ozone is a common subject of sensitivity analysis studies with atmospheric chemistry models, as in \cite{Constantin2014} and \cite{Christian2018}, since it has a high impact on air quality. We chose 2 hours as the time interval because we found that the surrogate from \cite{Kelp2020} was most accurate at this time scale, with MSE$(f)/\var{\Y} \approx 0.06$. 

Since we did not have access to a working implementation of the CBM-Z model, we used a dataset of 80,000 samples provided by \cite{Kelp2020} that was independent of the data used for training and validating their surrogate. The initial concentrations and meteorological inputs were sampled from independent uniform distributions with ranges outlined in \cite{Kelp2020}. In particular, the dataset consists of 625 i.i.d. batches of samples, where each batch is a full Latin hypercube with 128 points. Thus, we used sample sizes that were multiples of 128 and applied the CLT to the means within each batch for deriving confidence intervals. 

We again compare floodgate to the confidence intervals given by the surrogate-based SPF estimator $\Shat_j^f$ and using the error bounds from \cite{Panin2021} for every input. Since the implementation of the CBM-Z model used by \cite{Kelp2020} to train their surrogate was proprietary and not available to us, we were not able to evaluate $\fstar$ ourselves and thus could not implement the non-surrogate baseline for these experiments. As discussed in Section~\ref{sec: computational_savings}, the fact that floodgate can be applied to a pre-existing dataset is itself an advantage over the estimator $\Shat_j$.

As in Sections~\ref{sec: experiments} and \ref{sec: compare_application}, for a given computational budget $N$, floodgate uses the full set of $N$ samples from $\fstar$ to estimate each sensitivity index. \cite{Panin2021}'s bounds use all $N$ samples to estimate $\mathcal{E}$. Again, since $\Shat_j^f$ does not use any evaluations of $\fstar$, it has no dependence on $N$, so we choose to evaluate it with $n=N$ terms in the summation for each sensitivity index as well.

\subsection{Results}
\begin{figure}
\begin{center}
\includegraphics[width=\textwidth]{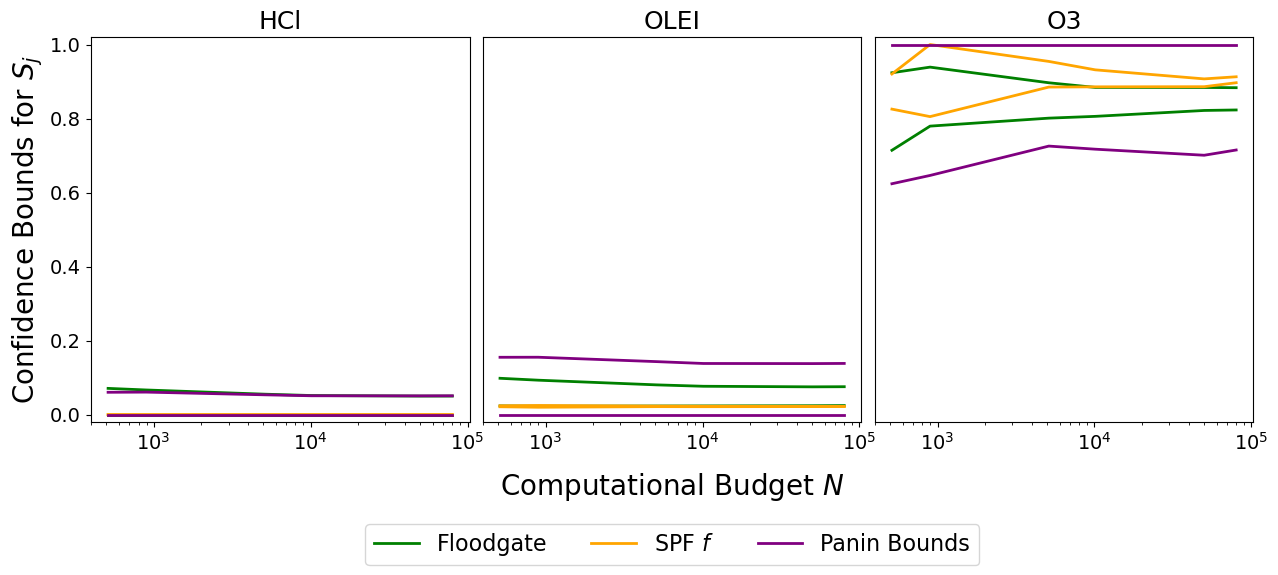}
\end{center}
\caption{95\% confidence intervals for a representative subset of sensitivity indices for the CBM-Z model using floodgate, the surrogate-based SPF estimator, and \cite{Panin2021}'s bounds.}
\label{fig: CBMZ_intervals}
\end{figure}

Figure~\ref{fig: CBMZ_intervals} shows the confidence intervals obtained by floodgate, the surrogate-based SPF estimator, and \cite{Panin2021}'s bounds for a representative subset of the inputs (the plots for the full set of inputs can be found in Appendix~\ref{apx: cbmz}). In this case, since we have only a small dataset that does not contain samples in the paired form required for an SPF estimator, we cannot obtain sufficiently precise estimates of the ground truth sensitivities to add the horizontal red lines to the plots or to calculate coverage.

For most of the inputs, the estimated sensitivities were very close to 0, and their plots look almost identical to that for HCl. This is because the green and purple curves are essentially the same when $\Shat_j^f\approx 0$, since both give lower bounds of 0, and the error bound in \eqref{eqn: panin} will be simply $\mathcal{E}^2$, which is the same as the width of our intervals. However, when $\Shat_j^f$ is even slightly above 0, the gap between the green and purple upper bounds becomes quite noticeable, as in the plot for OLEI. For O3, which naturally has the highest sensitivity of all the inputs, \cite{Panin2021}'s confidence interval is roughly twice as wide as ours for $N=512$, and it is nearly five times wider than ours for $N=80000$. The intervals derived using the surrogate-only method (orange curves) converges to a value outside our interval entirely, which again demonstrates that this na\"{i}ve method can give high confidence for an incorrect result.

%% file: conclusion.tex
\section{Conclusion}
We present a novel method, floodgate, for conducting statistically rigorous sensitivity analysis using surrogate models to achieve substantial computational speedups relative to existing methods. Floodgate provides asymptotically valid confidence intervals whose accuracy directly improves with the quality of the surrogate used. Since all of our theoretical results are very general, the method can be applied to any computational model with any surrogate, allowing users to take full advantage of arbitrary domain knowledge and state-of-the-art machine learning models, regardless of their complexity, to achieve results that are as accurate as possible. Furthermore, the confidence intervals provide rigorous quantification of the uncertainty of surrogate-based estimates, informing the user as to when they can and cannot be trusted.

We highlight some of the advantages of floodgate compared to existing work empirically through simulations with the relatively simple Hymod hydrological model in Section~\ref{sec: simulations} and in a more realistic application to the CBM-Z meteorological model in Section~\ref{sec: application}. These results validate the fact that having a high-quality surrogate allows us to obtain very tight bounds, and when this is not the case, our confidence intervals account for the surrogate's inaccuracy and still provide valid coverage. We compare floodgate both theoretically and empirically to similar results for statistically valid surrogate-based inference from \cite{Panin2021}, and we show that the confidence intervals we provide are significantly narrower while still maintaining coverage. 

More broadly, we see the future implications of our work in this paper as aiding in the general goal of being able to study complex, expensive models through their less-expensive surrogates without sacrificing statistical guarantees. Computational models are used in several high-stakes settings, aiding in important scientific discoveries, shaping engineering designs, and making forecasts or simulations that inform high-impact decisions; thus it is crucial to have both an accurate understanding of the input-output relationships in these models, as well as to be conscious of the uncertainty in the studies we perform. This principle extends beyond just sensitivity analysis, applying to all forms of inference on the models' features and outputs in which one might be interested. 

%% file: acknowledgements.tex
\section{Acknowledgements}
The authors would like to thank Lu Zhang for sharing early drafts of her PhD thesis that our work built upon and Makoto Kelp for generously providing code and data necessary for the CBM-Z application. 
M. A. and L.J. were partially supported by a CAREER grant from the National Science Foundation (Grant \#DMS2045981).

%% file: appendix/lower_bound_numerator.tex
\subsection{Numerator of $\ell(f)$}
\label{apx: lj_numerator}

From \eqref{eqn: u_i} and \eqref{eqn: ell_i}, we can write
$$\ell(f) = \frac{\E{(\Y-\fZ(\Xcomp))^2} - \E{(\Y-f(X,Z))^2}}{\var{\Y}}.$$
When $\var{\Y}=0$, we have $\Y \eqas \E{\Y} =: a$. In this case, the numerator of $\ell(f)$ simplifies to
\begin{align}
& \E{(a-\fZ(\Xcomp))^2} - \E{(a-f(X,Z))^2} \notag\\
= \; & a^2 - 2a\E{\fZ(\Xcomp)} + \E{\fZ^2(\Xcomp)} - \left(a^2 - 2a\E{f(X,Z)} + \E{f^2(X,Z)}\right) \notag\\
= \; &\E{\fZ^2(\Xcomp)} - \E{f^2(X,Z)} - 2a\left(\E{\fZ(\Xcomp)} - \E{f(X,Z)}\right) \notag\\
= \; &-\E{\fZ^2(\Xcomp)} + 2\E{\fZ(\Xcomp)f(X,Z)} - \E{f^2(X,Z)} - 2a\left(\E{f(X,Z)} - \E{f(X,Z)}\right) \notag\\
= \; &-\E{(f(X,Z) - \fZ(\Xcomp))^2}, \label{eqn: l_numerator}
\end{align}
where the third equality follows from the law of iterated expectations:
    $$\E{\fZ(\Xcomp)} = \E{\fZfull} = \E{f(X,Z)}$$
    and
    $$\E{\fZ(\Xcomp)f(X,Z)} = \E{\E{\fZ(\Xcomp)f(X,Z) | \Xcomp}} = \E{\fZ(\Xcomp)\E{f(X,Z) | \Xcomp}} = \E{\fZ^2(\Xcomp)}.$$
Note that the term in \eqref{eqn: l_numerator} is always non-positive. In particular, it will be 0 when $\var{f(X,Z) | \Xcomp} \eqas 0$, meaning that $\ell(f)=0$ by definition, and it will be negative when $\E{\var{f(X,Z) | \Xcomp}} > 0$, meaning that $\ell(f)$ can take the value $-\infty$.

%% file: appendix/bounds.tex
\subsection{Proof of Lemma \ref{lem: bounds}}
\label{prf: bounds}

\textbf{Lemma.} \textit{For any $\fstar$ and $f$ such that $\ell(f)$ and $u(f)$ exist,}
\begin{equation}
    \ell(f) \le \ell(\fstar) = S = u(\fstar) \le u(f).\notag
\end{equation}

\noindent \emph{Proof.} Recall the definition of the function $\fZ(\Xcomp) = \fZfull$ from Section \ref{sec: bounds}, and similarly define
$$\fZstar(\Xcomp) = \fZfullstar.$$
We can rewrite the expressions for $u(f)$ and $S$ given in \eqref{eqn: u_i} and \eqref{eqn: Si} as 
$$u(f) = \frac{\E{\left(\Y-\fZ(\Xcomp)\right)^2}}{\var{\Y}} \qquad \text{and} \qquad S = \frac{\E{\left(\Y-\fZstar(\Xcomp)\right)^2}}{\var{\Y}}.$$
When $\var{\Y}=0$, $S=0$ by definition, $u(f)$ can take the values 0 or $\infty$, and $\ell(f)$ can take the values $0$ or $-\infty$, so the inequality holds.

When $\var{\Y} > 0$, in order to prove that $u(f)\ge S$, it suffices to show $D_1 \defeq \E{\left(\Y-\fZ(\Xcomp)\right)^2} - \E{\left(\Y-\fZstar(\Xcomp)\right)^2}\ge 0$. With some algebraic manipulation, we have
\begin{align}
	D_1 &= \E{\left(\Y-\fZstar(\Xcomp) + \fZstar(\Xcomp) - \fZ(\Xcomp)\right)^2} - \E{\left(\Y-\fZstar(\Xcomp)\right)^2} \notag\\
	&= \E{\left(\Y-\fZstar(\Xcomp)\right)^2} + 2\E{\left(\Y-\fZstar(\Xcomp)\right)\left(\fZstar(\Xcomp) - \fZ(\Xcomp)\right)} \notag\\
	&\qquad \qquad + \E{\left(\fZstar(\Xcomp) - \fZ(\Xcomp)\right)^2} - \E{\left(\Y-\fZstar(\Xcomp)\right)^2} \notag\\
	&= 2\E{\left(\Y-\fZstar(\Xcomp)\right)\left(\fZstar(\Xcomp) - \fZ(\Xcomp)\right)} + \E{\left(\fZstar(\Xcomp) - \fZ(\Xcomp)\right)^2} \label{eqn: expanded_1} 
\end{align}
We can expand the expectation in the first term in \eqref{eqn: expanded_1} to get
\begin{align*}
	\label{eqn: u_cross_term}
	&\mathbb{E}\big[(\Y-\fZstar(\Xcomp))\left(\fZstar(\Xcomp) - \fZ(\Xcomp)\right)\big] \\
	= \; &\E{\Y\left(\fZstar(\Xcomp) - \fZ(\Xcomp)\right)} - \E{\fZstar(\Xcomp)\left(\fZstar(\Xcomp) - \fZ(\Xcomp)\right)} \\
	= \; &\E{\E{\Y(\fZstar(\Xcomp)-\fZ(\Xcomp))\mid \Xcomp}} - \E{\fZstar(\Xcomp)(\fZstar(\Xcomp)-\fZ(\Xcomp))} \\
	= \; &\E{\E{\Y\mid \Xcomp}(\fZstar(\Xcomp)-\fZ(\Xcomp))} - \E{\fZstar(\Xcomp)(\fZstar(\Xcomp)-\fZ(\Xcomp))} \\
	= \; &\E{\fZstar(\Xcomp)(\fZstar(\Xcomp)-\fZ(\Xcomp))} - \E{\fZstar(\Xcomp)(\fZstar(\Xcomp)-\fZ(\Xcomp))} \\
	= \; &0.
\end{align*} 
Therefore, by plugging this into \eqref{eqn: expanded_1}, we have
\begin{equation}
\label{eqn: D_1}
	D_1 = \E{\left(\fZstar(\Xcomp) - \fZ(\Xcomp)\right)^2} \ge 0.
\end{equation}
Now, we consider the lower bound $\ell(f)$ from \eqref{eqn: ell_i}. Again, to prove $\ell(f) \le S$, it suffices to show $D_2 \defeq \left(\E{\left(\Y-\fZ(\Xcomp)\right)^2} - \E{\left(\Y-f(X,Z)\right)^2}\right) - \E{\left(\Y-\fZstar(\Xcomp)\right)^2} \le 0$. Note that 
\begin{equation}
	\label{eqn: D_2D_1}
	D_2 = D_1 - \E{\left(\Y-f(X,Z)\right)^2}.
\end{equation}
Since we found $D_1$ in \eqref{eqn: D_1}, we just need to deal with this second term. Define the functions 
\rev{\begin{align*}
	\hj(X,Z) &= f(X,Z) - \fZ(\Xcomp)\\
	\hstar(X,Z) &= \Y - \fZstar(\Xcomp).
\end{align*}}
Thus, we can rewrite 
\rev{\begin{align}
	&\; \E{\left(\Y-f(X,Z)\right)^2} \notag\\
	= & \;\E{\left(\hstar(X,Z)+\fZstar(\Xcomp) - \hj(X,Z)-\fZ(\Xcomp)\right)^2} \notag\\
	= & \; \E{\left(\hstar(X,Z)-\hj(X,Z)\right)^2} + 2\E{\left(\hstar(X,Z)-\hj(X,Z)\right)\left(\fZstar(\Xcomp)-\fZ(\Xcomp)\right)} \notag \\
	&\quad + \E{\left(\fZstar(\Xcomp)-\fZ(\Xcomp)\right)^2}. \label{eqn: mse_term}
\end{align}}
From \eqref{eqn: D_1}, \eqref{eqn: D_2D_1}, and \eqref{eqn: mse_term}, we have
\rev{\begin{equation}
	D_2 = -\E{\left(\hstar(X,Z)-\hj(X,Z)\right)^2} - 2\E{\left(\hstar(X,Z)-\hj(X,Z)\right)\left(\fZstar(\Xcomp)-\fZ(\Xcomp)\right)}. \label{eqn: D_2_step2}
\end{equation}}
Simplifying the expectation in the second term in \eqref{eqn: D_2_step2}, we have
\rev{\begin{align*}
	\E{\left(\hstar(X,Z)-\hj(X,Z)\right)\left(\fZstar(\Xcomp)-\fZ(\Xcomp)\right)} &= \E{\E{\left(\hstar(X,Z)-\hj(X,Z)\right)\left(\fZstar(\Xcomp)-\fZ(\Xcomp)\right)\mid \Xcomp}} \\
	&= \E{\E{\left(\hstar(X,Z)-\hj(X,Z)\right)\mid\Xcomp}\left(\fZstar(\Xcomp)-\fZ(\Xcomp)\right)} \\
	&= 0,
\end{align*}}
where the last equality holds because $\E{\hstar(X,Z)\mid\Xcomp} \eqas \E{\hj(X,Z)\mid\Xcomp} \eqas 0$ by definition. Finally, plugging this into \eqref{eqn: D_2_step2}, we have
\rev{\begin{equation}
	\label{eqn: D_2_final}
	D_2 = -\E{\left(\hstar(X,Z)-\hj(X,Z)\right)^2} \le 0. 
\end{equation}}
Finally, it is clear from \eqref{eqn: D_1} and \eqref{eqn: D_2_final} that
\begin{equation}
	\ell(\fstar) = S = u(\fstar).\tag*{\qed}
\end{equation}

%% file: appendix/estimator.tex
\subsection{Proof of Lemma \ref{lem: unbiased}}
\label{prf: unbiased}

\textbf{Lemma.} \textit{For $K\ge 1$, given a set of i.i.d. original samples $\{\Xj\}_{i=1}^n$ and modified samples $\tilde{X}_i^{(1)},\dots, \tilde{X}_i^{(K)} \sim \Pcond$ such that $\Xnew \indp \Xij | \Xcompj$ for each $i,k$, then for any $\fstar, f: \Rd\to\R$, the quantity $\Wj$ defined in \eqref{eqn: Wij} satisfies:}
\begin{equation*}
    \E{\Wj} = \textnormal{MSE}(\fZ).
\end{equation*}

\noindent \emph{Proof.} We start by defining the term
$$\Fij := \frac{1}{K}\sum_{k=1}^K f(\Xnew, \Xcompj),$$
so we can rewrite the expression for $\Wj$ from \eqref{eqn: Wij} as
$$\Wj = \left(\Yj - \Fij\right)^2 - \frac{1}{K+1}\left(f\Xj - \Fij\right)^2,$$
Note that 
\begin{equation}
	\label{eqn: lin_exp}
	\E{\Wj} = \E{\left(\Y - \Fij\right)^2} - \frac{1}{K+1}\E{\left(f(X,Z) - \Fij\right)^2},
\end{equation}
by linearity of expectation, so we can examine each term on the right hand side of \eqref{eqn: lin_exp} separately. From here on, we drop the index $i$ on random variables within expectations since samples are i.i.d. Expanding the first term, we have
\begin{equation}
	\label{eqn: expand_t1}
	\E{\left(\Y - \Fj\right)^2} = \E{\Ysquared} - 2\E{\Y \Fj} + \E{\Fj^2}.
\end{equation}
Simplifying the expectation in the second term of \eqref{eqn: expand_t1}, we get
\begin{align}
    \E{\Y\Fj} &= \frac{1}{K}\E{\Y\sumK f(\Xtildek,\Xcomp)} \notag\\
    &= \frac{1}{K}\sumK \E{\Y f(\Xtildek,\Xcomp)} \notag\\
    &= \E{\Y f(\rev{\Xtilde^{(1)}},\Xcomp)} \notag\\
    &= \E{\E{\fstar(X, \Xcomp) f(\rev{\Xtilde^{(1)}},\Xcomp) \mid \Xcomp}} \notag\\
    &= \E{\E{\fstar(X, \Xcomp) \mid \Xcomp}\E{f(\rev{\Xtilde^{(1)}},\Xcomp) \mid \Xcomp}} \notag\\
    &= \E{\Y\E{f(X,Z)\mid\Xcomp}} \notag\\
    &= \E{\Y\fZ(\Xcomp)}, \label{eqn: fstarF}
\end{align}
where the third and sixth equalities use the fact that $\{X,\tilde{X}^{(1)},\dots, \tilde{X}^{(K)}\}$ are all exchangeable, the fourth and sixth lines apply the law of iterated expectations, and the fifth line follows from the conditional independence of $X$ and $\rev{\Xtilde^{(1)}}$. 

Simplifying the expectation in the third term of \eqref{eqn: expand_t1}, we get
\begin{align}
    \E{\Fj^2} &= \frac{1}{K^2}\E{\left(\sumK f(\Xtildek, \Xcomp)\right)^2} \notag\\
    &= \frac{1}{K^2}\E{\sumK f^2(\Xtildek,\Xcomp) + 2\sumK\sum_{t=k+1}^K f(\Xtildek,\Xcomp)f(\tilde{X}^{(t)},\Xcomp)} \notag\\
    &= \frac{1}{K^2}\sumK\E{f^2(\Xtildek,\Xcomp)} + \frac{2}{K^2}\sumK\sum_{t=k+1}^K \E{f(\Xtildek,\Xcomp)f(\tilde{X}^{(t)},\Xcomp)} \notag\\
    &= \frac{1}{K}\E{f^2(X,Z)} + \frac{K-1}{K}\E{f(X,\Xcomp)f(\rev{\Xtilde^{(1)}},\Xcomp)} \notag\\
    &= \frac{1}{K}\E{f^2(X,Z)} + \frac{K-1}{K} \E{\E{f(X,\Xcomp) f(\rev{\Xtilde^{(1)}},\Xcomp)\mid\Xcomp}} \notag\\
    &= \frac{1}{K}\E{f^2(X,Z)} + \frac{K-1}{K} \E{\E{f(X,\Xcomp)\mid\Xcomp} \E{f(\rev{\Xtilde^{(1)}},\Xcomp)\mid\Xcomp}} \notag\\
    &= \frac{1}{K}\E{f^2(X,Z)} +\frac{K-1}{K}\E{(\E{f(X,Z) \mid\Xcomp})^2} \notag\\
    &= \frac{1}{K}\E{f^2(X,Z)} + \frac{K-1}{K}\E{\fZ^2(\Xcomp)}, \label{eqn: Fsquared}
\end{align}
where the fourth and seventh equalities follow from the exchangeability of $\{X,\tilde{X}^{(1)},\dots, \tilde{X}^{(K)}\}$, the fifth line follows from the law of iterated expectations, and the sixth line follows from the conditional independence of $X$ and $\rev{\Xtilde^{(1)}}$.

Substituting \eqref{eqn: fstarF} and \eqref{eqn: Fsquared} into \eqref{eqn: expand_t1}, we get
\begin{align}
    &\E{(\Y-\Fj)^2} \notag\\
    = \; &\E{\Ysquared} - 2\E{\Y\fZ(\Xcomp)} + \frac{1}{K}\E{f^2(X,Z)} + \frac{K-1}{K}\E{\fZ^2(\Xcomp)} \notag\\
    = \; &\E{\Ysquared} - 2\E{\Y\fZ(\Xcomp)} + \E{\fZ^2(\Xcomp)} + \frac{1}{K}\left(\E{f^2(X,Z)} - \E{\fZ^2(\Xcomp)}\right) \notag\\
    = \; &\E{(\Y-\fZ(\Xcomp))^2} + \frac{1}{K}\E{(f(X,Z)-\fZ(\Xcomp))^2}, \label{eqn: simplify_t1}
\end{align}
where in the second term in the last line, we use the fact that the cross term $\E{f(X,Z)\fZ(\Xcomp)} = \E{\fZ^2(\Xcomp)}$ by the law of iterated expectations.

Now we can simplify the expectation in the second term of \eqref{eqn: lin_exp}. Expanding, we have
\begin{align}
	&\E{\left(f(X,Z)-\Fj\right)^2} \notag\\
	= \; &\E{\left(f(X,Z)-\frac{1}{K}\sum_{k=1}^K f(\Xtildek, \Xcomp)\right)^2} \notag \\
	= \; &\E{f^2(X,Z)} - \frac{2}{K}\sum_{k=1}^K \E{f(X,Z)f(\Xtildek, \Xcomp)} + \E{\left(\frac{1}{K}\sum_{k=1}^K f(\Xtildek, \Xcomp)\right)^2} \notag \\
	= \; &\E{f^2(X,Z)} - 2\E{f(X,Z)f(\rev{\Xtilde^{(1)}}, \Xcomp)} + \frac{1}{K^2}\E{\left(\sum_{k=1}^K f(\Xtildek, \Xcomp)\right)^2}. \label{eqn: substitute_F_t2}
\end{align}
We have already found the expectation \rev{of} the last term here in \eqref{eqn: Fsquared}. The middle term is very similar to \eqref{eqn: fstarF}:
\begin{align}
    \E{f(X,Z)f(\rev{\Xtilde^{(1)}},\Xcomp)} &= \E{\E{f(X, \Xcomp) f(\rev{\Xtilde^{(1)}},\Xcomp) \mid \Xcomp}} \notag\\
    &= \E{\E{f(X, \Xcomp) \mid \Xcomp}\E{f(\rev{\Xtilde^{(1)}},\Xcomp) \mid \Xcomp}} \notag\\
    &= \E{\fZ^2(\Xcomp)}. \label{eqn: fF}
\end{align}
Substituting \eqref{eqn: Fsquared} and \eqref{eqn: fF} into \eqref{eqn: substitute_F_t2}, we get
\begin{align}
    \E{(f(X,Z)-\Fj)^2} &= \E{f^2(X,Z)} - 2\E{\fZ^2(\Xcomp)} + \frac{1}{K}\E{f^2(X,Z)} + \frac{K-1}{K}\E{\fZ^2(\Xcomp)} \notag\\
    &= \frac{K+1}{K}\left(\E{f^2(X,Z)} - \E{\fZ^2(\Xcomp)}\right) \notag\\
    &= \frac{K+1}{K}\E{(f(X,Z) - \fZ(\Xcomp))^2}, \label{eqn: simplify_t2}
\end{align}
where the last line follows for the same reason as in \eqref{eqn: simplify_t1}.

Finally, plugging \eqref{eqn: simplify_t1} and \eqref{eqn: simplify_t2} into \eqref{eqn: lin_exp}, we have
\begin{align}
	\E{\Wj} = \; &\E{(\Y - \fZ(\Xcomp))^2} + \frac{1}{K}\E{(f(X,Z)-\fZ(\Xcomp))^2} \notag\\
	&- \; \frac{1}{K+1}\left(\frac{K+1}{K}\right)\E{(f(X,Z)-\fZ(\Xcomp))^2}\notag\\
	= \; &\E{(\Y - \fZ(\Xcomp))^2}\notag\\
	= \; &\text{MSE}(\fZ). \tag*{\qed}
\end{align}

%% file: appendix/validity.tex
\subsection{Proof of Theorem \ref{thm: validity}}
\label{prf: validity}

\textbf{Theorem.} \textit{For i.i.d. samples $\{\Xj\}_{i=1}^n$, any computational model $\fstar$ and surrogate $f:\Rd\to\R$, and any $\alpha\in (0,1)$, if $\E{{\fstar}^4(X,Z)}, \E{f^4(X,Z)} < \infty$, then the bounds $\Lower$ and  $\Upper$ output by Algorithm~\ref{alg: conf} satisfy}
$$\liminf_{n\rightarrow\infty}\pr{\Lower \le S \le \Upper} \ge 1 - \alpha.$$

\noindent \emph{Proof.} We first handle the case where $\var{\Y}=0$. Since this implies that $\bar{V}\eqas 0$, Algorithm~\ref{alg: conf} will return $[0,1]$ with probability 1. Since $S=0$ by definition when $\var{\Y}=0$, the floodgate interval has 100\% coverage.

Now we consider the case where $\var{\Y} > 0$. The first step in this proof is to show that $\Mjbar$, $\bar{M}$ and $\bar{V}$ are all asymptotically normal and consistent for MSE$(\fZ)$, MSE$(f)$, and $\var{\Y}$, respectively. Since these estimators are sample means of i.i.d. terms that we have already established are unbiased for their corresponding estimands, it suffices to show that $\var{\Wj}, \var{\Zj}, \var{V_i} < \infty$, or equivalently that their second moments are finite, in order to apply the CLT. As in the previous proof, we will drop the index $i$ on random variables withing expectations, since samples are i.i.d. 

Starting with $M$, we have
\begin{align*}
	\E{M^2}	= \; &\E{(\Y-f(X,Z))^4}\\
	= \; &\E{{\fstar}^4(X,Z)} - 4\E{{\fstar}^3(X,Z)f(X,Z)} + 6\E{\Ysquared f^2(X,Z)} \\
	&- \; 4\E{\Y f^3(X,Z)} + \E{f^4(X,Z)}\\
	< \; &\infty,
\end{align*}
where the final inequality holds because the first and last terms are finite by assumption, and the middle three terms can be bounded by H\"{o}lder's inequality:
\begin{align*}
    \left|\E{{\fstar}^3(X,Z)f(X,Z)}\right| &\le \E{\left|{\fstar}^3(X,Z)f(X,Z)\right|}\\
    &\le \left(\E{\left|{\fstar}^3(X,Z)\right|^\frac{4}{3}}\right)^\frac{3}{4} \left(\E{|f(X,Z)|^4}\right)^\frac{1}{4}\\
    &= \left(\E{{\fstar}^4(X,Z)}\right)^\frac{3}{4} \left(\E{f^4(X,Z)}\right)^\frac{1}{4} \\
    &< \infty\\
    0 \le \E{{\fstar}^2(X,Z)f^2(X,Z)} &\le \left(\E{\left|{\fstar}^2(X,Z)\right|^2}\right)^\frac{1}{2} \left(\E{|f^2(X,Z)|^2}\right)^\frac{1}{2}\\
    &= \left(\E{{\fstar}^4(X,Z)}\right)^\frac{1}{2} \left(\E{f^4(X,Z)}\right)^\frac{1}{2} \\
    &< \infty\\
    \left|\E{\fstar(X,Z)f^3(X,Z)}\right| &\le \E{\left|\fstar(X,Z)f^3(X,Z)\right|}\\
    &\le \left(\E{\left|\fstar(X,Z)\right|^4}\right)^\frac{1}{4} \left(\E{|f^3(X,Z)|^\frac{4}{3}}\right)^\frac{3}{4}\\
    &= \left(\E{{\fstar}^4(X,Z)}\right)^\frac{1}{4} \left(\E{f^4(X,Z)}\right)^\frac{3}{4} \\
    &< \infty.
\end{align*}
Next, we consider $\W$. Define
\begin{align}
	T_1 &\defeq \left(\Y - \Fj\right)^2 \label{eqn: T_1}\\
	\text{and } \; T_2 &\defeq \left(f(X,Z) - \Fj\right)^2, \label{eqn: T_2}
\end{align}
so that $\W = T_1 + \frac{1}{K+1}T_2$. Then,
\begin{align}
	\E{\W^2} &= \E{T_1^2} - \frac{2}{K+1}\E{T_1T_2} + \frac{1}{(K+1)^2}\E{T_2^2} \notag\\
	&\le \E{T_1^2} + \frac{1}{(K+1)^2}\E{T_2^2}, \label{eqn: Wi_sq}
\end{align}
since $T_1,T_2\ge 0$ almost surely. We first consider $T_1$:
\begin{align}
	&\E{T_1^2} \notag\\
	= \; &\E{\left(\Y - \Fj\right)^4} \notag\\
	= \; &\E{{\fstar}^4(X,Z)} - 4\E{{\fstar}^3(X,Z)\Fj} + 6\E{{\fstar}^2(X,Z)\Fj^2} - 4\E{\Y\Fj^3} + \E{\Fj^4}. \label{eqn: T1_sq}
\end{align}
The first term is finite by assumption, so $\E{\Fj^4}<\infty$ is sufficient to show $\E{T_1^2}<\infty$, since the middle three terms can all be bounded by H\"{o}lder's inequality. Substituting in the expression for $\Fj$, we have
\begin{align}
	\E{\Fj^4} = \frac{1}{K^4}\E{\left(\sumK f(\Xtildek,\Xcomp)\right)^4}. \label{eqn: Fi_4th}
\end{align}
While we spare the messy algebra here, expanding the right-hand side of \eqref{eqn: Fi_4th} yields a sum of the terms $\E{f^4(X,Z)}$, $\E{f^3(\Xtilde^{(1)},\Xcomp)f(\Xtilde^{(2)},\Xcomp)}$, $\E{f^2(\Xtilde^{(1)},\Xcomp)f^2(\Xtilde^{(2)},\Xcomp)}$,\\ $\E{f^2(\Xtilde^{(1)},\Xcomp)f(\Xtilde^{(2)},\Xcomp)f(\Xtilde^{(3)},\Xcomp)}$, and $\E{f(\Xtilde^{(1)},\Xcomp)f(\Xtilde^{(2)},\Xcomp)f(\Xtilde^{(3)},\Xcomp)f(\Xtilde^{(4)},\Xcomp)}$ with finite coefficients. The first of these expectations is finite by assumption, which also implies the existence of the second expectation by applying H\"{o}lder's inequality with $p=\frac{4}{3}$ and of the last three expectations by repeated application of Cauchy--Schwarz.
Now we consider $T_2$:
\begin{align}
	&\E{T_2^2} \notag\\
	= \; &\E{\left(f(X,Z) - \Fj\right)^4} \notag\\
	= \; &\E{f^4(X,Z)} - 4\E{f^3(X,Z)\Fj} + 6\E{f(X,Z)^2\Fj^2} - \E{f(X,Z)\Fj^3} + \E{\Fj^4}. \label{eqn: T2_sq}
\end{align}
As in \eqref{eqn: T1_sq}, the first term is finite by assumption, and since we already already established $\E{\Fj^4} < \infty$, we can bound the middle three terms using H\"{o}lder's inequality, and thus $\E{T_2^2} < \infty$. By \eqref{eqn: Wi_sq}, \eqref{eqn: T1_sq}, and \eqref{eqn: T2_sq}, we have $\E{\W^2} < \infty$. 

Finally, $\E{V^2} < \infty$ is immediate given $\E{{\fstar}^4(X,Z)}<\infty$.

Now, by the multivariate CLT,
$$\sqrt{n}\Bigg(\begin{pmatrix} \Mjbar \\ \bar{M} \\ \bar{V}\end{pmatrix} - \begin{pmatrix} \text{MSE}(\fZ) \\ \text{MSE}(f) \\ \var{\Y} \end{pmatrix}\Bigg) \convd\mathcal{N}\left(0,\Sigma\right),$$
where 
$$\Sigma = \begin{bmatrix} 
\Sigma_{11} & \Sigma_{12} & \Sigma_{13} \\ 
\Sigma_{12} & \Sigma_{22} & \Sigma_{23} \\ 
\Sigma_{13} & \Sigma_{23} & \Sigma_{33}\end{bmatrix} 
\defeq \begin{bmatrix}
\var{\W} & \cov{\W}{M} & \cov{\W}{V} \\
\cov{\W}{M} & \var{M} & \cov{M}{V} \\
\cov{\W}{V} & \cov{M}{V} & \var{V} \\
\end{bmatrix}.$$

Since we assume that $\var{\Y} > 0$, we can apply the multivariate delta method to get
\begin{align}
	\frac{\sqrt{n}}{\sigma_\ell}\left(\frac{\Mjbar - \bar{M}}{\bar{V}} - \frac{\text{MSE}(\fZ) - \text{MSE}(f)}{\var{\Y}}\right) = \frac{\sqrt{n}}{\sigma_\ell}\left(\ellhat - \ell(f)\right) &\convd \mathcal{N}(0,1), \label{eqn: ellhat_normal}\\
	\frac{\sqrt{n}}{\sigma_u}\left(\frac{\Mjbar}{\bar{V}} - \frac{\text{MSE}(\fZ)}{\var{\Y}}\right) = \frac{\sqrt{n}}{\sigma_u}\left(\uhat - u(f)\right) &\convd \mathcal{N}(0,1) \label{eqn: uhat_normal}
\end{align}
where
\begin{align*}
	\sigma_\ell^2 &\defeq \frac{1}{\left(\var{\Y}\right)^2}\Bigg(\Sigma_{11} + \Sigma_{22} + \left(\frac{\text{MSE}(\fZ)-\text{MSE}(f)}{\var{\Y}}\right)^2\Sigma_{33} - 2\Sigma_{12} \\
	&\qquad\qquad\qquad\qquad\qquad \;- 2\frac{\text{MSE}(\fZ) - \text{MSE}(f)}{\var{\Y}}\left(\Sigma_{23} -\Sigma_{13}\right)\Bigg)\\
	\sigma_u^2 &\defeq \frac{1}{\left(\var{\Y}\right)^2}\left(\Sigma_{11} - 2\frac{\text{MSE}(\fZ)}{\var{\Y}}\Sigma_{13} + \frac{\text{MSE}(\fZ)}{\left(\var{\Y}\right)^2}\Sigma_{33}\right).
\end{align*}

By Slutsky's Theorem, we can replace $\sigma_\ell$ and $\sigma_u$ in \eqref{eqn: ellhat_normal} and \eqref{eqn: uhat_normal} with their consistent estimators $s_\ell$ and $s_u$ defined in Algorithm \ref{alg: conf}, and the same results hold. 

\rev{
Note that
\begin{equation*}
    \Lower = \left\{\begin{array}{cl} 0 & \text{if }\bar{V}=0, \\
    0 & \text{if }\ellhat - \frac{z_{\alpha/2} s_\ell}{\sqrt{n}}<0, \\
    \ellhat - \frac{z_{\alpha/2} s_\ell}{\sqrt{n}} & \text{else},
    \end{array}\right.
\end{equation*}
and since $\Lower\le S$ in either of the first two cases (since $S\ge 0$ by definition), we have that
\begin{align}
\label{eqn: eq_prob_lower}
\pr{\Lower \le S} &\ge \pr{\ellhat - \frac{z_{\alpha/2} s_\ell}{\sqrt{n}} \le S}, \end{align}
where we say that the event $\{\ellhat - \frac{z_{\alpha/2} s_\ell}{\sqrt{n}} \le S\}$ does not occur when $\ellhat - \frac{z_{\alpha/2} s_\ell}{\sqrt{n}}$ is undefined (due to $\bar{V}=0$). By the same argument, 
\begin{align}
\label{eqn: eq_prob_upper}
\pr{\Upper \ge S} &\ge \pr{\uhat + \frac{z_{\alpha/2} s_u}{\sqrt{n}} \ge S}. \end{align}
Thus, by \eqref{eqn: eq_prob_lower} and \eqref{eqn: ellhat_normal}, 
\begin{align*}
    \liminf_{n\to\infty}\pr{\Lower \le S} &\ge \liminf_{n\to\infty}\pr{\ellhat - \frac{z_{\alpha/2} s_\ell}{\sqrt{n}} \le S} \\
    &\ge \liminf_{n\to\infty}\pr{\ellhat - \frac{z_{\alpha/2} s_\ell}{\sqrt{n}} \le \ell(f)} = 1 - \frac{\alpha}{2},
\end{align*}
and by \eqref{eqn: eq_prob_upper} and \eqref{eqn: uhat_normal}, 
\begin{align*}
    \liminf_{n\to\infty}\pr{\Upper \le S} &\ge \liminf_{n\to\infty}\pr{\uhat + \frac{z_{\alpha/2} s_u}{\sqrt{n}} \ge S} \\
    &\ge \liminf_{n\to\infty}\pr{\uhat + \frac{z_{\alpha/2} s_u}{\sqrt{n}} \ge u(f)} = 1 - \frac{\alpha}{2}.
\end{align*}
}

A simple union bound gives us the final result:
\begin{equation}
	\liminf_{n\rightarrow\infty}\pr{\Lower \le S \le \Upper} \ge 1 - \alpha. \tag*{\qed}
\end{equation}

%% file: appendix/width.tex
\subsection{Proof of Theorem \ref{thm: width}}
\label{prf: width}

\textbf{Theorem.} \textit{Under the same assumptions as in Theorem~{\ref{thm: validity}} and the additional assumption that $\var{\Y}>0$, the bounds $\Lower$ and  $\Upper$ output by Algorithm~\ref{alg: conf} satisfy}
$$\Upper - \Lower \rev{\le} \frac{\textnormal{MSE}(f)}{\var{\Y}} + O_p\left(n^{-1/2}\right).$$

\noindent \emph{Proof.} Note that since $\bar{V}\convp\var{\Y} > 0$, $\pr{\bar{V} = 0}\to 0$. When $\bar{V} > 0$,
\begin{align}
	\left(\Upper - \Lower\right) - \frac{\textnormal{MSE}(f)}{\var{\Y}} &\rev{\le} \left(\frac{\Mjbar}{\bar{V}} + \frac{z_{\alpha/2}s_u}{\sqrt{n}}\right) - \left(\frac{\Mjbar - \bar{M}}{\bar{V}} - \frac{z_{\alpha/2}s_\ell}{\sqrt{n}}\right) - \frac{\text{MSE}(f)}{\var{\Y}}\notag \\
	&= \left(\frac{\bar{M}}{\bar{V}} - \frac{\text{MSE}(f)}{\var{\Y}}\right) + \frac{z_{\alpha/2}(s_u+s_\ell)}{\sqrt{n}}. \label{eqn: O_nhalf}
\end{align}
The reason for the inequality in the first line is that we could have $\Lower = 0$ or $\Upper = 1$, since we clip the bounds. Using the intermediate results from the proof of Theorem \ref{thm: validity} and our assumption that $\var{\Y}>0$, we can apply the multivariate delta method to establish that
$$\sqrt{n}\left(\frac{\bar{M}}{\bar{V}} - \frac{\text{MSE}(f)}{\var{\Y}}\right) \convd \mathcal{N}(0,\sigma^2),$$
for some $\sigma^2 < \infty$. This implies that the first term in \eqref{eqn: O_nhalf} is $O_p(n^{-1/2})$. Since $s_u\convp\sigma_u$ and $s_\ell\convp\sigma_\ell$, $s_u$ and $s_\ell$ are $O_p(1)$, meaning that the second term in \eqref{eqn: O_nhalf} is also $O_p(n^{-1/2})$. This establishes the desired result.
\begin{equation}
\text{ } \tag*{\qed}
\end{equation}

%% file: appendix/hymod.tex
\subsection{Description of Hymod Model}
\label{apx: hymod}

Hymod  is a conceptual rainfall-runoff model based on the probability-distributed soil storage capacity principle \citep{Sheikholeslami2020}. It simulates daily streamflow given precipitation and potential evapotranspiration as forcing data, where both the inputs and outputs are time series. The model consists of a soil moisture module and a routing module composed of three linear quick flow reservoirs and one linear slow flow reservoir. The outputted streamflow is given by the sum of quick and slow flow generation \citep{Herman2013}.

\begin{table}[H]
\begin{center}
\begin{tabular}{c | c c c c}
Input	& Description							& Units				& Min	& Max\\
\hline
Sm			& maximum soil moisture 					& mm					& 0		& 400\\
beta			& exponent in the soil moisture routine	& -					& 0		& 2\\
alfa 		& partition coefficient 					& -					& 0		& 1\\
Rs   		& slow reservoir coefficient 			& $\text{day}^{-1}$	& 0		& 0.1\\
Rf   		& fast reservoir coefficient 			& $\text{day}^{-1}$	& 0.1	& 1\\
\end{tabular}
\caption{Hymod inputs \citep{Pianosi2016}}
\label{tab: hymod_parameters}
\end{center}
\end{table}

%% file: appendix/CBMZ.tex
\subsection{Full CBM-Z Sensitivity Indices}
\label{apx: cbmz}

\begin{figure}[H]
\begin{center}
\includegraphics[height=0.85\textheight]{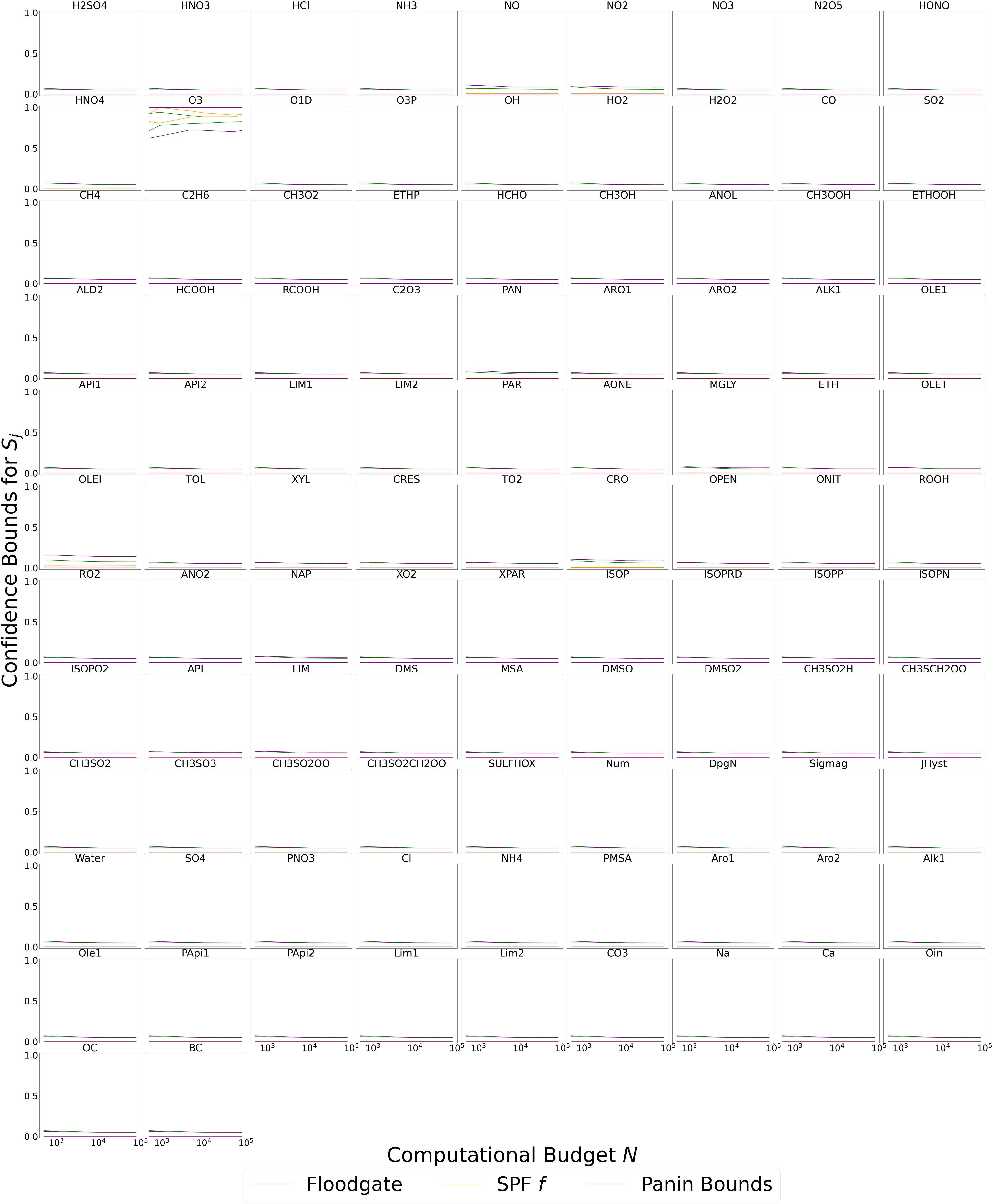}
\end{center}
\caption{95\% confidence intervals for all sensitivity indices for the CBM-Z model using floodgate, the surrogate-based SPF estimator, and \cite{Panin2021}'s bounds.}
\label{fig: CBMZ_full}
\end{figure}